%% file: main.tex
\documentclass[submission, Phys]{SciPost}

\input{incl_settings.tex} 
\input{incl_shortcuts.tex}

\graphicspath{{./figs_new/}}

\begin{document}

\begin{center}{\Large \textbf{
Unbinning global LHC analyses
}}\end{center}

\begin{center}
Henning Bahl\textsuperscript{1},
Tilman Plehn\textsuperscript{1,2}, and
Nikita Schmal\textsuperscript{1}
\end{center}

\begin{center}
{\bf 1} Institut f\"ur Theoretische Physik, Universit\"at Heidelberg, Germany\\
{\bf 2} Interdisciplinary Center for Scientific Computing (IWR), Universit\"at Heidelberg, Germany\\
\end{center}

\begin{center}
\today
\end{center}

\section*{Abstract}
{\bf 
    Neural simulation-based inference has been shown to outperform traditional, histogram-based inference in numerous phenomenological and experimental studies at the LHC. So far, these analyses have focused on individual processes. We study the combination of four different di-boson processes in terms of the Standard Model Effective Field Theory. Our results demonstrate how neural simulation-based inference also wins over traditional methods for more global LHC analyses.
}

\vspace{10pt}
\noindent\rule{\textwidth}{1pt}
\tableofcontents\thispagestyle{fancy}
\noindent\rule{\textwidth}{1pt}
\vspace{10pt}

\clearpage
%%%%%%%%%%%%%%%%%%%%%%%%%%%%%%%%%%%%%%%%%%%%%%%%%%%%%%

\clearpage
%%%%%%%%%%%%%%%%%%%%%%%%%%%%%%%%%%%%%%%%%%%%%%%%%%%%%%
\section{Introduction}

One of the biggest upcoming challenges in particle physics is to exploit the full potential of the current and future LHC datasets. Since the Higgs boson discovery, the LHC has been turned into a precision hadron collider, but extracting the full information from high-dimensional event data is becoming an increasingly problematic bottleneck. Traditional inference methods rely on, at most, few-dimensional histograms as summary statistics. This reduction of dimensionality and the binning result in a loss of information. Modern machine learning (ML) methods can overcome this bottleneck~\cite{Butter:2022rso,Plehn:2022ftl}. In particular, they can be used to extract statistically powerful unbinned likelihood ratios~\cite{Cranmer:2019eaq}.

Besides the ML-based matrix-element method~\cite{Butter:2022vkj,Heimel:2023mvw}, neural simulation-based inference (SBI)~\cite{Brehmer:2018eca,Brehmer:2018kdj,Brehmer:2018hga,Brehmer:2019xox,Chatterjee:2021nms,Chatterjee:2022oco,Kong:2022rnd,Schofbeck:2024zjo,Bahl:2024meb,Silva:2025hzo} is a proven method for likelihood ratio estimation. The advantages of SBI for LHC processes have been demonstrated not only in a variety of phenomenological studies~\cite{Brehmer:2019gmn,Barman:2021yfh,Bahl:2021dnc,Chatterjee:2021nms,Chatterjee:2022oco,GomezAmbrosio:2022mpm,Barrue:2023ysk,Mastandrea:2024irf,Schofbeck:2024zjo,Ghosh:2025fma,Benato:2025rgo}, but recently also in the first experimental SBI analysis at the LHC~\cite{ATLAS:2024jry,ATLAS:2025clx}. While SBI already improves the sensitivity to individual theory parameters, it has even greater potential if multiple theory parameters are to be inferred. Since SBI does not rely on lower-dimensional summary statistics, it is much more effective in disentangling different theory parameters than binned histograms.

A key aspect of modern LHC physics is the combination of many processes into a global analysis, to answer whether LHC data is explained by the Standard Model everywhere. The standard framework for such global analyses is the Standard Model Effective Field Theory (SMEFT)~\cite{Degrande:2012wf,Brivio:2017vri}. Using SMEFT, numerous global LHC analyses have demonstrated that much stronger constraints on fundamental parameters can be obtained from the combination of measurements and processes~\cite{Butter:2016cvz,daSilvaAlmeida:2018iqo,Biekotter:2018ohn,Ellis:2018gqa,Almeida:2021asy,ATLAS:2022xyx,Giani:2023gfq,Elmer:2023wtr,Heimel:2024drk,deBlas:2025xhe,CMS:2025ugn}. These analyses are currently based on total rate or binned (differential) rate measurements. While it is methodologically clear how to integrate SBI into a global SMEFT analysis~\cite{GomezAmbrosio:2022mpm}, it is not clear to what level the advantages of SBI persist for such global analyses. 

We aim to quantify this effect for four different di-boson processes: $WW$, $WZ$, $WH$, and $ZH$ production. These processes are highly relevant in global SMEFT analyses of the combined electroweak and Higgs sector~\cite{Butter:2016cvz,Brivio:2022hrb}: $WW$ and $WZ$ production constrain the resulting anomalous triple gauge couplings and purely electroweak operators, while $WH$ and $ZH$ are sensitive to Higgs–gauge interactions and fermionic contact terms. Focusing on a subset of SMEFT operators, to which the di-boson processes are dominantly sensitive, we compare the sensitivity of the SBI and histogram methods in detail.

The paper is structured as follows. In Sec.~\ref{sec:methods}, we describe the SBI methods used. Section~\ref{sec:physics} provides an overview of the considered processes and SMEFT operators. We present results in Sec.~\ref{sec:results}. Conclusions can be found in Sec.~\ref{sec:conclusions}.

%%%%%%%%%%%%%%%%%%%%%%%%%%%%%%%%%%%%%%%%%%%%%%%%%%%%%%
\section{Methodology}
\label{sec:methods}

We start by describing our methodology including learning the likelihood ratio using the derivative-learning approach, backgrounds, fractional smearing, and limit setting.

%%%%%%%%%%%%%%%%%%%%%%%%%%
\subsection{Learning the likelihood ratio}

At parton level, the likelihood of the observables $z_p$ given a parameter value $\theta$ is directly related to the differential cross section,
\begin{align}
  p(z_p|\theta) = \frac{1}{\sigma(\theta)}\frac{\d\sigma(z_p|\theta)}{\d z_p} \; .
\end{align}
The differential cross-section can be computed via the matrix element $\mat$ and the parton distribution functions $f_{1,2}$,
\begin{align}
  d\sigma(z_p|\theta) 
  = (2\pi)^4 \int dx_1dx_2 d\Phi \frac{f_1(x_1,Q^2)f_2(x_2,Q^2)}{2x_1 x_2 s}|\mat(z_p|\theta)|^2 \;.
\end{align}
Assuming that the theory parameters $\theta$ only affect the hard scattering element, the reconstruction-level likelihood is given by~\cite{Brehmer:2018eca}
\begin{align}
  p(x|\theta) 
  &= \int dz_d dz_s dz_p p(x|z_d) p(z_d|z_s) p(z_s|z_p) p(z_p|\theta)  \notag \\
  &=\int dz_p p(x|z_p) p(z_p|\theta)\;. 
\label{eq:pxtheta_int}
\end{align}
Here, $p(z_s|z_p)$ encodes the parton shower and hadronisation; $p(z_d|z_s)$, the detector response; and, $p(x|x_d)$, the reconstruction of the reconstruction-level observables $x$. 

The integral in Eq.\eqref{eq:pxtheta_int} is not tractable, implying that the $p(x|\theta)$ cannot be directly computed. We can, however, use neural networks to learn the likelihood ratio,
\begin{align}
    r(x|\theta,\theta_0)\equiv \frac{p(x|\theta)}{p(x|\theta_0)}\;,
\end{align}
where $\theta_0$ is a reference hyptothesis --- e.g., the SM. This can be achieved for example using the loss function
\begin{align}
    \loss &= 
    \XXLangle \left[ r(z_p|\theta,\theta_0)- r_\varphi(x|\theta,\theta_0) \right]^2 \XXRangle_{x,z_p \sim p(x|z_p)p(z_p|\theta);\theta\sim q(\theta)}  \; ,
    \label{eq:MSE_ratio_loss}
\end{align}
where $r_\varphi$ is the network output and $r(z_p|\theta,\theta_0)$ is the tractable parton-level likelihood ratio. The squared difference is averaged over paired parton-level and reconstruction-level Monte-Carlo samples with the theory parameters sampled from a prior $q(\theta)$.

Learning the likelihood ratio conditioned on the theory parameters is difficult and often numerically unstable. This can be avoided by exploiting the known dependencies of the likelihood ratio on the theory parameters. It is particularly useful for SMEFT, whose Lagrangian has the structure
\begin{align}
    \lag_\text{SMEFT} 
    = \lag_\text{SM} + \sum_{i}\frac{c_i}{\Lambda^2} \; O_i 
    \equiv \lag_\text{SM} + \sum_{i}\theta_i \; O_i \;.
\end{align}
Here, the $c_i$'s are the Wilson coefficients and $\Lambda$ is the new physics scale up to which the EFT is valid. We truncate the expansion in $1/\Lambda^2$ after the first order and chose to infer $\theta_i\equiv c_i/\Lambda^2$. Expanding the squared matrix element to second order in the $\theta$, the differential cross-section ratio
\begin{align}
   R(x|\theta,\theta_0) 
   \equiv \frac{d\sigma(x|\theta)/dx}{d\sigma(x|\theta_0)/dx} 
   = \frac{\sigma(\theta)p(x|\theta)}{\sigma(\theta_0)p(x|\theta_0)} \; ,
\end{align}
can be expanded up to the same second order
\begin{align}
       R(x|\theta,\theta_0) 
        &= 1 + (\theta-\theta_0)_i R_i(x) 
            + (\theta-\theta_0)_i (\theta-\theta_0)_j R_{ij}(x) \notag \\
    R_i(x) &\equiv \frac{\partial}{\partial \theta_i} R(x|\theta,\theta_0)\bigg|_{\theta = \theta_0} \notag \\
    R_{ij}(x) &\equiv \frac{\partial^2}{\partial\theta_i \partial\theta_j} R(x|\theta,\theta_0)\bigg|_{\theta = \theta_0} \, .
\label{eq:Rexp}
\end{align}
As in Eq.\eqref{eq:MSE_ratio_loss}, we can learn these coefficients independently of each other using as targets the parton-level ratios~\cite{Chatterjee:2021nms,Chatterjee:2022oco}
\begin{alignat}{3}
    & R_i(z_p) &&\equiv \frac{\partial}{\partial\theta_i}\frac{d\sigma(z_p|\theta)/dz_p}{d\sigma(z_p|\theta_0)/dz_p}\Bigg|_{\theta=\theta_0}
    &&= \frac{\partial_{\theta_i} |\mat(z_p|\theta)|^2}{|\mat(z_p|\theta_0)|^2}
    \Bigg|_{\theta_0} \notag \\
    & R_{ij}(z_p) &&\equiv \frac{\partial^2}{\partial\theta_i\partial\theta_j}\frac{d\sigma(z_p|\theta)/dz_p}{d\sigma(z_p|\theta_0)/dz_p}\Bigg|_{\theta=\theta_0}
    &&= \frac{\partial_{\theta_i}\partial_{\theta_j} |\mat(z_p|\theta)|^2}{|\mat(z_p|\theta)|^2}
    \Bigg|_{\theta_0} \; .
\end{alignat}
The neural network estimators $R_\varphi$ approximate the true ratios $R_i(x)$ and $R_{ij}(x)$,
\begin{align}
 R_{\varphi,i}(x) &\approx R_i(x) \notag \\
 R_{\varphi,ij}(x) &\approx R_{ij}(x) \; .
\end{align}
As training data, events sampled from $\theta_0$ --- i.e., from the SM --- are used. The likelihood ratio is then obtained via
\begin{align}
    r(x|\theta,\theta_0) &=  \frac{\sigma(\theta_0)}{\sigma(\theta)}\frac{d\sigma(x|\theta)/dx}{d\sigma(x|\theta_0)/dx} = 
    \frac{\sigma(\theta_0)}{\sigma(\theta)} R(x|\theta,\theta_0) \notag\\
    &\approx 1 + (\theta-\theta_0)_i R_{\varphi,i}(x) 
            + (\theta-\theta_0)_i (\theta-\theta_0)_j R_{\varphi,ij}(x)\;.
\end{align}
We refer to this approach as derivative learning.

Alternatively, the likelihood ratio can be learned at specific benchmark points. The theory parameter dependence is then recovered by morphing the likelihood between these benchmark points~\cite{Brehmer:2018eca,Bahl:2024meb}. In this work, we only use derivative learning as we previously found it to be more stable if the phase-space regions populated by the SM and BSM hypotheses have significant overlap~\cite{Bahl:2024meb}.

%%%%%%%%%%%%%%%%%%%%%%%%%%
\subsection{Backgrounds}
\label{sec:backgrounds}

In the presence of background, we can further split up the squared matrix element into a signal, a background, and an interference component,
\begin{align}
    |\mat(z_p|\theta)|^2 = |\mat_\sig(z_p|\theta)|^2 + 2\text{Re}\left[\mat_\sig(z_p|\theta) \mat_\bkg^*(z_p)\right] + |\mat_\bkg(z_p)|^2\;.
\end{align}
Notably, the background matrix element is independent of the theory parameters $\theta$.
For the processes considered in this work, the interference contribution is negligible --- either due to different initial/final states or to the small Higgs boson width. In this case, the differential cross section can be written as the sum
\begin{align}
    d\sigma(x|\theta) = d\sigma_\sig(x|\theta) + d\sigma_\bkg(x)\; .
\end{align}
Correspondingly, the likelihood and likelihood ratio split into 
\begin{align}
    p(x|\theta) &=  \frac{\sigma_\text{sig}(\theta)}{\sigma_\text{sig}(\theta) + \sigma_\text{bkg}}p_\text{sig}(x|\theta) + \frac{\sigma_\text{bkg}}{\sigma_\text{sig}(\theta) + \sigma_\text{bkg}} p_\bkg(x) \notag \\
    r(x|\theta,\theta_0) 
    &=\frac{\sigma_\sig(\theta_0) + \sigma_\bkg}{\sigma_\sig(\theta) + \sigma_\bkg} \dfrac{1 + \dfrac{\sigma_\sig(\theta_0)}{\sigma_\bkg}\omega(x|\theta_0)R(x|\theta,\theta_0)}{1 + \dfrac{\sigma_\sig(\theta_0)}{\sigma_\bkg}\omega(x|\theta_0)} \notag \\
    &\text{with} \qquad 
    \omega(x|\theta_0) \equiv \frac{p_\sig(x|\theta_0)}{p_\bkg(x)}\;.
\end{align}
The likelihood ratio can be extracted through a signal--background classifier, whose classifier score $D$ converges towards~\cite{Cranmer:2015bka,Mastandrea:2024irf}
\begin{align}
    D_\text{opt}(x) = \frac{p_\sig(x|\theta_0)}{p_\sig(x|\theta_0) + p_\bkg(x))} = \frac{\omega(x)}{1 + \omega(x)} 
    \qquad \Leftrightarrow \qquad 
    \omega(x) = \frac{D_\text{opt}(x)}{1 - D_\text{opt}(x)}\;.
\end{align}
The first derivative of the summed log-likelihood ratio is required to vanish at $\theta = \theta_0$. This can be derived analytically:
\begin{align}
    &\frac{\partial}{\partial\theta_i}\left[\int dx\, p(x|\theta_0) \log r(x|\theta,\theta_0) \right]_{\theta_i = \theta_{0,i}} \notag \\
    &= \int \d x\, p(x|\theta_0) \frac{R_j(x)}{1 + \frac{\sigma_\bkg}{\sigma_\sig(\theta_0)}\frac{1}{\omega(x|\theta_0)}} - \frac{\sigma_{\sig,j}(\theta_0)}{\sigma_\sig(\theta_0)+\sigma_\bkg} \notag\\
    &= \frac{\sigma_\sig(\theta_0)}{\sigma_\sig(\theta_0)+\sigma_\bkg}\int \d x p_\sig(x|\theta_0)\frac{\partial}{\partial \theta_j} \frac{\d\sigma_\sig(x|\theta)}{\d\sigma_\sig(x|\theta_0)} - \frac{\sigma_{\sig,j}(\theta_0)}{\sigma_\sig(\theta_0)+\sigma_\bkg} \notag \\
    &= \frac{1}{\sigma_\sig(\theta_0)+\sigma_\bkg}\frac{\partial}{\partial \theta_j} \int \d x  \frac{\d\sigma_\sig(x|\theta)}{\d x} - \frac{\sigma_{\sig,j}(\theta_0)}{\sigma_\sig(\theta_0)+\sigma_\bkg}\notag\\
    &= 0 \;, \label{eq:logr_1st_derivative}
\end{align}
where we introduced
\begin{align}
    \sigma_{\sig,j}(\theta_0) = \frac{\partial}{\partial\theta_j} \sigma_\sig(\theta)\bigg|_{\theta = \theta_0}\;.
\end{align}
If the classifier and the differential cross-section ratios are not learned perfectly or if the data samples are not large enough, the numerical cancellation between the first and second terms above might not be perfect. The likelihood will then not have its minimum at $\theta_0$, even using a large dataset generated for $\theta = \theta_0$. To avoid this numerical issue, we set
\begin{align}
\sigma_\text{sig,j} =  \left(\sigma_\sig(\theta_0) + \sigma_\bkg\right) \mean{\frac{R_{\varphi,j}(x_i)}{1 + \frac{\sigma_\bkg}{\sigma_\sig(\theta_0)}\frac{1}{\omega(x_i|\theta_0)}} }_{x\sim p(x|\theta_0)}\;,
\end{align}
when evaluating the kinematic part of the likelihood. This extends the procedure proposed in Ref.~\cite{Bahl:2024meb} for the background-free case.

%%%%%%%%%%%%%%%%%%%%%%%%%%
\subsection{Fractional smearing}
\label{sec:frac_smear}

Another numerical problem arises from outlier events associated with significantly larger derivatives $R_{i}$ or $R_{ij}$ than the bulk of the distribution. These events are passed through the parton shower and detector simulation only once, and the learned estimators can be biased. As proposed in Ref.~\cite{Bahl:2024meb}, we use fractional smearing to avoid this issue. We pass each event with a large derivative through the parton shower and detector simulation $n$ times and assign each resulting event a weight $1/n$, where $n$ is chosen based on the size of the derivatives. These weights are then incorporated into the MSE loss. For the example of likelihood ratio regression for a fixed $\theta$, this reads
\begin{align}
    \loss 
    &= \XXLangle \left[ r(z_p|\theta,\theta_0)- r_\varphi(x|\theta) \right]^2 \XXRangle_{x,z_p \sim p(x|z_p)p(z_p|\theta)}  \notag \\
    &= \sum_i^N \; w_i \left[ r(z_{p,i}|\theta,\theta_0)- r_\varphi(x_i|\theta) \right]^2\;.
\label{eq:loss_fs}
\end{align}
%
%For further details, we refer to Ref.~\cite{Bahl:2024meb}.

%%%%%%%%%%%%%%%%%%%%%%%%%%
\subsection{Limit setting and empirical coverage}
\label{sec:limit_setting_and_coverage}

The full likelihood for a set of $n$ events $\{x\}$ is given by
\begin{align}
    p_{\text{full}}(\{x\}|\theta) = \text{Pois}(n | L\sigma(\theta)) \prod_i p(x_i | \theta) 
    \qquad \text{with} \qquad 
    \text{Pois}(k|\lambda) = e^{-\lambda}\frac{\lambda^k }{k!} \; .
\end{align}
Given the luminosity $L$, $\text{Pois}(n | L\sigma(\theta))$ is the total rate likelihood. The unbinned kinematic likelihood for each event is $p(x_i|\theta)$~\cite{Arratia:2021otl}. The corresponding likelihood ratio reads
\begin{align}
    \log r_\text{full}(\{x\}|\theta, \theta_0) = \log\frac{\text{Pois}(n | L\sigma(\theta))}{\text{Pois}(n | L\sigma(\theta_0))} + \sum_i\log r(x_i|\theta,\theta_0)\;.
\end{align}
To derive expected limits for $N_\text{exp}$ events, we replace the sum by
\begin{align}
    \sum_i \log r(x_i | \theta, \theta_0) 
    \; \to \; 
    \frac{N_\text{exp}}{N_\text{MC}} \Langle \log r(x_i | \theta, \theta_0) \Rangle_{x_i\in \{x\}_\text{MC}} \ ,
\end{align}
to exploit the full Monte-Carlo sample of size $N_\text{MC}$. For real observed data, the sum can be directly evaluated. In the presence of background, we can split this into
\begin{align}
     \frac{N_\text{exp}}{N_\text{MC}} \Langle \log r(x_i | \theta, \theta_0) \Rangle_{x_i\in \{x\}_\text{MC}} 
     \; \to \; & \frac{N_\text{bkg,exp}}{N_\text{bkg,MC}} \Langle \log r(x_i | \theta, \theta_0) \Rangle_{x_i\in \{x\}_\text{MC,bkg}} \notag\\
     & + \frac{N_\text{sig,exp}}{N_\text{sig,MC}} \Langle \log r(x_i | \theta,\theta_0) \Rangle_{x_i\in \{x\}_\text{MC,sig}}
\end{align}
using separate samples $\{x\}_\text{bkg}$ and $\{x\}_\text{sig}$ for background and signal, respectively. This is valid since background and signal samples are statistically independent. For multiple independent background processes, further splits are possible.

After deriving the full likelihood, the test statistic is given by
\begin{align}
    q(\theta) 
    &= -2 \log r_\text{full}(\{x\}|\theta, \hat\theta) \notag \\
    &= -2\left(\log r_\text{full}(\{x\}|\theta,\theta_0) - \log r_\text{full}(\{x\}|\hat\theta,\theta_0)\right)\;,
\end{align}
where $\hat\theta$ is the minimum of the likelihood. We approximate it by using
\begin{align}
    \hat\theta = \underset{\theta}{\text{argmax}}\log r_\text{full}(\{x\}|\theta,\theta_1) ; .
\end{align}
Based on Wilk's theorem, the distribution $p(q(\theta)|\theta)$ converges towards a chi-squared distribution for sufficiently many events. Based on this, we calculate the $p$-value for a parameter point $\theta$ via
\begin{align}
    p_\theta = \int_{q_\text{obs}(\theta)}^{\infty}dq\, p(q(\theta)|\theta) = 1 - F_{\chi^2}(q_\text{obs}(\theta)|k)\;.
\end{align}
It gives the confidence with which one can reject the parameter point $\theta$, where $q_\text{obs}(\theta)$ is the observed value of $q(\theta)$ for the sample $\{x\}$. $F_{\chi^2}(y|k)$ is the cumulative chi-squared distribution function with $k$ degrees of freedom. The $\gamma$ confidence region is defined by all $\theta$ values for which $p_\theta < \gamma$.

As a cross-check for the learned likelihood ratio $r_\varphi(x|\theta,\theta_0)$, we take $n$ samples for a given $\theta$ and evaluate for what fraction of samples the true $\theta$ lies within a given $\gamma$ confidence region, defining the coverage
\begin{align}
    c_\gamma \equiv \left\langle \mathbb{1}\left(p_{\theta_0}(\{x\}) > 1-\gamma\right) \right\rangle_{\{x\}_{1,...,n}}\;.
\end{align}
Here, we use the indicator function $\mathbb{1}$ which is equal to one if the expression in the brackets is fulfilled and zero otherwise. If the fraction is higher than the nominal confidence level, $c_\gamma > \gamma$, our learned likelihood is conservative or underconfident. Inversely, if it is lower the learned likelihood is overconfident.

%%%%%%%%%%%%%%%%%%%%%%%%%%%%%%%%%%%%%%%%%%%%%%%%%%%%%%
\section{Processes, operators, and training setup}
\label{sec:physics}

In our analysis we focus on four di-boson processes --- $WZ$, $WW$, $ZH$, and $WH$ production --- and a selected subset of SMEFT operators.

%%%%%%%%%%%%%%%%%%%%%%%%%%
\subsection{SMEFT operators}

%--------------------------------
\begin{table}[b!]
    \centering
    \begin{small}
    \begin{tabular}{ll|ll}
    \toprule
        \makebox[2cm][l]{operator} & definition & \makebox[2cm][l]{operator} & definition \\
        \midrule
        $\OHD$ & $(\Phi^\dagger D^\mu \Phi)^*(\Phi^\dagger D_\mu \Phi)$ &
        $\OHq$ & $\sum_{i= 1,2}  (\Phi^\dagger i \overset{\leftrightarrow}{D^a}_\mu \Phi)(\bar q_i\sigma^a \gamma^\mu q_i)$\\
        % $\OHbox$ & $(\Phi^\dagger\Phi)\Box(\Phi^\dagger\Phi)$ \\
        % \midrule
        $\OHB$ & $\Phi^\dagger\Phi B_{\mu\nu} B^{\mu\nu}$ & $\OHW$ & $\Phi^\dagger\Phi W_{\mu\nu}^a W^{\mu\nu a}$ \\
        $\OHWB$ & $\Phi^\dagger\sigma^a\Phi W_{\mu\nu}^a B^{\mu\nu}$ & $\OWWW$ & $\epsilon^{abc}W_\mu^{\nu a}W_\nu^{\rho b}W_\rho^{\mu c}$ \\
        % \midrule
        \bottomrule
    \end{tabular}
    \end{small}
    \caption{Dimension-6 SMEFT operators considered in our analysis.}
    \label{tab:Operators}
\end{table}
%--------------------------------

Throughout our analysis, we neglect flavor- or $\mathcal{CP}$-violating operators for simplicity. We, moreover, restrict ourselves to dimension-six operators. A list of all operators considered is given in Table~\ref{tab:Operators}. In this analysis we ignore the fact that they are also constrained by electroweak precision observables~\cite{Ellis:2018gqa,Biekotter:2018ohn,deBlas:2025xhe}. We list the relevant SMEFT Feynman rules in App.~\ref{app:SMEFT_Feynman_rules}.

%--------------------------------
\begin{table}[t]
    \centering
    \begin{small}
    \begin{tabular}{c|ccccccc}
        \toprule
        process & \OHD & \OHW & \OHB & \OHWB & \OHq & \OWWW \\
        \midrule
        $WZ$ & \checkmark & & & \checkmark & \checkmark & \checkmark \\
        $WW$ & \checkmark & & & \checkmark & \checkmark & \checkmark\\
        $ZH$ & \checkmark & \checkmark & \checkmark & \checkmark & \checkmark & \\
        $WH$ & \checkmark & \checkmark & & & \checkmark & \\
        \bottomrule
    \end{tabular}
    \end{small}
    \caption{Operators contributing to the different di-boson processes.}
    \label{tab:WC_contr}
\end{table}
%--------------------------------

Table~\ref{tab:WC_contr} details which operators affect which process. While $WH$ production is only affected by three of the considered operators, five operators contribute to $ZH$ production. At first sight, this larger set of relevant operators significantly increases the complexity of inferring the likelihood ratio. In practice, however, the expansion performed in Eq.\eqref{eq:Rexp} allows for a straightforward inclusion of more operators.

%%%%%%%%%%%%%%%%%%%%%%%%%%
\subsection{Di-boson processes}
\label{sec:di-boson_procs}

In our analysis we focus on four di-boson production processes: $qq\to WZ$, $qq\to WW$, $qq\to WH$, and $qq\to ZH$ production. We neglect the subleading loop-induced $gg\to ZH$ contribution to $ZH$ production. Regarding the boson decay channels, we consider the leptonic decay channels of the $W$ and $Z$ boson and the $H\to b\bar b$ decay channel. The resulting signatures are
\begin{equation}
\begin{aligned}
    qq' &\to W^\pm Z \to \ell^\pm\ell^+\ell^- + \ETmiss \\
    q\bar q &\to W^\pm W^\mp \to \ell^+\ell^- + \ETmiss \\
    q\bar q &\to ZH \to \ell^+\ell^- + b\bar b \\
    qq' &\to W^\pm H \to \ell^\pm + b\bar b + \ETmiss \; .
\end{aligned}
\end{equation}
%

%%%%
\subsubsection*{Pre-selection cuts and backgrounds}

%--------------------------------
%{
%\renewcommand{\arraystretch}{1.3}
\begin{table}[b!]
\centering
\begin{small}
\begin{tabular}{l|c}
\toprule
\textbf{processes} & \textbf{pre-selection cuts} \\
\midrule
jet/lepton &
$p_T^\ell > 15\ \gev$,\;
$p_T^j > 20\ \gev$,\;
$|\eta^i| < 2.5$ \; with  $i \in \{\ell,j\}$ \\
\midrule
$WZ$ &
$\ETmiss > 45\ \gev$,\;
$p_T^{\ell W} > 20\ \gev$,\;
$m_T^W > 30\ \gev$, \;
$81.2 < m_Z^{\ell\ell} < 101.2\ \gev$,\; $N_\ell = 3$ \\[1mm]
%\midrule
$WW$ &
$\ETmiss > 45\ \gev$,\;
$m_{\ell\ell} > 15\ \gev$,\;
$|m_{\ell\ell} - m_Z| > 15\ \gev$, \; 
$N_\ell = 2$,\; $N_j = 0$ \\[1mm]
%\midrule
$WH$ and $ZH$ &
$p_T^b > 35\ \gev$,\;
$p_T^j < 30\ \gev$,\;
$80 < m_{bb} < 160\ \gev$, \; 
$R_{bb},\ R_{b\ell},\ R_{bj},\ R_{\ell j} > 0.4$ \\[1mm]
%\midrule
$WH$ only &
$\ETmiss > 25\ \gev$,\; $N_\ell = 1$,\; $N_b = 2$ \\[1mm]
%\midrule
$ZH$ only &
$N_\ell = N_b = 2$ \\
\bottomrule
\end{tabular}
\end{small}
\caption{Summary of analysis selection cuts.}
\label{tab:cuts}
\end{table}
%}
%--------------------------------

We impose a series of pre-selection cuts, inspired by experimental analyses~\cite{ATLAS:2019rob,ATLAS:2019bsc,ATLAS:2025dhf,ATLAS:2022ooq,CMS:2021fyk}. These cuts are chosen to suppress backgrounds. All cuts are listed in Tab.~\ref{tab:cuts}, where the object definition cuts for leptons $\ell$ and jets $j$ are universal. The employed high-level observables are
\begin{itemize}
    % \item $p_T$: transverse momentum,
    % \item $\eta$: pseudorapidity,
    % \item $\ETmiss$: missing transverse energy,
    \item $p_T^{\ell W}$: reconstructed transverse momentum of the lepton originating from the $W$ boson decay,
    \item $m_T^W$: transverse mass of the $W$ boson,
    \item $m_Z^{\ell\ell}$: invariant mass of the lepton pair originating from the $Z$ boson decay,
    \item $m_{\ell\ell}$: di-lepton invariant mass,
    \item $m_{bb}$: invariant mass of the bottom pair.
    % \item $R_{ij}$: angular distance between the $i$ and $j$ particles.
\end{itemize}
For $WZ$ production, the main relevant backgrounds originate from $Z+\text{jets}$, $Z\gamma$, $t\bar t$, and $WW$ production with a fake lepton in the final state or --- without a fake lepton --- from $ZZ$, $t\bar t V$, or $VVV$ production, where $V$ is either a $W$ or $Z$ boson. In the signal region, the contribution of these backgrounds is about 20\%~\cite{ATLAS:2019bsc}. Since we expect the histogram-based and SBI approaches to be affected in a similar manner, we do not consider the backgrounds for $WZ$ production for our analysis. 

Similarly, we also do not take the backgrounds for $WW$ production into account. Here, the main background would be di-top production, which is, however, heavily suppressed by the applied jet veto. Further subleading backgrounds are Drell-Yan, $W + \text{jets}$, and $WZ$ production. In the signal region, the signal contribution is dominant with the background constituting about 35\% of the total yield~\cite{ATLAS:2019rob}.

For $ZH$ production, the dominant background is $Zb\bar b$ production. In the pre-selection region, the $Zb\bar b$ contribution is larger than the $ZH$ contribution. Therefore, we include this background in our analysis following the procedure outlined in Sec.~\ref{sec:backgrounds}. However, we neglect the SMEFT corrections to the background process, assuming that the trained signal--background classifier will suppress the contribution of the background processes to the final likelihood. The presence of additional Wilson coefficients like four-fermion operators could compensate for the impact of the considered operators. We do not include them since they do not affect the target di-boson processes. The cancellation for the $Zb\bar b$ process can, however, still be present in particular given that $Zb\bar b$ production is precisely measured with the results being well compatible with the SM expectations.

For $WH$ production, we include the three most relevant backgrounds, $t\bar t$ and $Wb\bar b$, where $t\bar t b\bar b$ is simulated as part of $t\bar t$. For the $Wb\bar b$ background, we again neglect the dependence on the considered Wilson coefficients based on the same considerations as for the $Zb\bar b$ background.

In more realistic analyses, the signal processes will not be perfectly separated. To handle this, the SBI analysis can be performed per signal region with the signal being compromised out of the different signal processes.

%%%%
\subsubsection*{Histogram observables}

\begin{table}[b!]
    \centering
    \begin{small}
    \begin{tabular}{l|lc}
        \toprule
        process & observable & binning \\
        \midrule
        $WW$ & $p_T^{\ell_1}$ & [0, 40, 50, 60, 70, 80, 90, 120, 140, 160, 180, 200, 300, 500]\\
        $WZ$ & $m_T^{WZ}$ & [0, 200, 400, 600, 800, 1000, 1500, 2500] \\
        $WH$ & $p_{T}^W$ & [0, 75, 150, 250, $\infty$]\\
        $ZH$ & $p_{T}^Z$ & [0, 75, 150, 250, $\infty$]\\
        \bottomrule
    \end{tabular}
    \end{small}
    \caption{Observables and bins used for the histogram limit setting.}
    \label{tab:hist_obs}
\end{table}

For the histogram limit setting, we have to choose specific observables and binning. Our selection shown in Tab.~\ref{tab:hist_obs} is based on existing experimental analysis for $WW$ and $WZ$ production~\cite{ATLAS:2019bsc,ATLAS:2025dhf}, as well as the simplified template cross-section (STXS) stage~1.2~\cite{LHCHiggsCrossSectionWorkingGroup:2016ypw,Andersen:2016qtm,Berger:2019wnu,Amoroso:2020lgh} for $WH$ and $ZH$ production. For the $WW$ and $WZ$ binning, $p_T^{\ell_1}$ is the $p_T$ of the leading lepton and $m_T^{WZ}$, the transverse mass of the $WZ$ system. The choice of observables and binnings can be improved to provide better sensitivity to our SMEFT operators~\cite{Brehmer:2019gmn}. We deliberately adopt the standard STXS binning, to get an indication of how much sensitivity can be gained by adopting an SBI approach in comparison to existing experimental analyses.

For the histogram limits, we derive the respective histograms for the parameter $\theta$ and $\theta_0$ and use them as input for a log-likelihood test. In the presence of background, we weight the histogram entries by the classifier output. This gives better sensitivity than placing a selection cut based on the classifier score.

%%%%
\subsubsection*{Event generation and dataset preparation}

For event generation, we employ \madgraph~3.5.0~\cite{Alwall:2011uj}. Generation is done at leading order, employing the \textsc{SMEFTatNLO}~\cite{Degrande:2020evl} \textsc{UFO} model. We approximate the effect of next-to-leading order corrections by applying a flat $K$-factor~\cite{FebresCordero:2009xzo,Czakon:2011xx,Gehrmann:2014fva,LHCHiggsCrossSectionWorkingGroup:2016ypw,Denner:2020eck,Buonocore:2022pqq}. The boson decays are simulated with \textsc{MadSpin}~\cite{Artoisenet:2012st}. For the parton shower, we use \pythia~8.306~\cite{Sjostrand:2014zea}; for the detector simulation, \delphes~3.5.0~\cite{deFavereau:2013fsa}; and, for the jet algorithm, \fastjet~3.3.4~\cite{Cacciari:2011ma}. We use the parton distribution function set \pdfset~\cite{Butterworth:2015oua} accessed via \lhapdf~\cite{Buckley:2014ana}. All events are generated for a center-of-mass energy of $13.6\,\tev$. Using this setup, we generate $\sim 10^6$ events for each signal and background process after applying the preselection cuts. For the signal channels, we further generate separate fractionally smeared datasets of size $\sim 2\cdot 10^6$.

Although the considered operators can modify the decays of the $Z$, $W$, and Higgs bosons, we neglect their effect, which is limited by the low momentum transfer in the decays. Most of the branching ratios are already known to agree with the SM prediction with high precision. The impact of large Wilson coefficients on these decays can, moreover, be compensated for by other Wilson coefficients, which we do not include in our analysis.

Finally, we neglect systematic uncertainties. These affect the non-kinematic as well as the kinematic part of the likelihood. The total rate contribution to the likelihood is the same for the SBI and histogram approaches. Given that SBI extracts more kinematic information, allowing to decorrelate the effects of systematic uncertainties from the theory parameters, we expect the systematic uncertainties to degrade the histogram limits more than the SBI limits. We refer to Refs.~\cite{Schofbeck:2024zjo,Benato:2025rgo} for a detailed treatment of systematic uncertainties in the SBI approach.

%%%%%%%%%%%%%%%%%%%%%%%%%%
\subsection{Neural network setup and training}

%%%%
\subsubsection*{Learning differential cross-section ratios} 

For learning the differential cross-section ratios $R_i$ and $R_{ij}$, we use a multi-layer perceptron (MLP) with five hidden layers, each consisting of 128 nodes. As activation function, we use the cumulative distribution function of the Gaussian distribution (GELU). As inputs, we use the four-vectors of all final-state particles, complemented by a selection of high-level features like the $VV$-invariant mass. Training is done for each process separately using 80\% of the fractionally smeared dataset for training and 20\% for validation. After training for 100--200 epochs using a cosine annealing scheduler with an initial learning rate of $10^{-4}$, we select the best model based on the validation loss.

By default, we regress all non-zero $R_i$ and $R_{ij}$ simultaneously with a single MLP having one output node per target summing the corresponding losses. Only for the $ZH$ and $WW$ processes, we learn $R_{\cHD}$ and $R_{\cHD^2}$ with seperate MLPs to improve the numerical stability. For the $WH$ process, we also exploit that \cHD only rescales the SM amplitude, as can be seen via the Feynman rules listed in App.~\ref{app:SMEFT_Feynman_rules}. Consequently, $R_{\cHD}$ and $R_{\cHD^2}$ are constants, and $R_{\cHD\cHq}\sim R_{\cHq}$ as well as $R_{\cHD\cHW} \sim R_{\cHW}$.

As discussed in Sec.~\ref{sec:limit_setting_and_coverage}, we checked the coverage of the learned likelihoods finding good agreement with the nominal confidence level. More detailed results are collected in App.~\ref{app:coverage}.

%%%%
\subsubsection*{Signal--background classifiers}

For the signal--background classification, we use a simple MLP with five layers, each containing 256 nodes. The output of each layer is passed through a leaky ReLU activation function. As inputs, we use four-vector entries of the involved particles as well as a selection of high-level observables like the $WH$ or $ZH$ invariant mass. The networks are trained on balanced signal and background datasets, to which no fractional smearing is applied. 

For $WH$ production, the background sample is obtained by mixing the samples of the different background processes according to their relative contribution to the phase-space region defined by the pre-selection cuts. 70\% of these datasets are used for training, while the remaining 30\% are used for the actual limit setting. For training, we use a scheduler which reduces the learning rate if the training loss has reached a plateau with an initial learning rate of $10^{-3}$, a decay factor of 0.1, and a patience of 10 epochs. We find very good performance after approximately 50 epochs. The achieved area-under-the-curve (AUC) values for the receiver-operating characteristic (ROC) curve are 0.81 and 0.84 for $WH$ and $ZH$ production, respectively.

%%%%%%%%%%%%%%%%%%%%%%%%%%%%%%%%%%%%%%%%%%%%%%%%%%%%%%
\section{Results}
\label{sec:results}

In this Section, we present the confidence limits for the di-boson processes introduced in Sec.~\ref{sec:physics} using the methods of Sec.~\ref{sec:methods}. We assume an integrated luminosity of $300\,\text{fb}^{-1}$, deriving expected limits based on the SM assumption. Throughout this section, we compare
\begin{enumerate} 
\item limits based on the known parton-level likelihood ratio for reference; 
\item limits based solely on the total rate as a lower sensitivity bound;
\item limits derived from reconstruction-level histograms; and
\item limits from the reconstruction-level SBI.
\end{enumerate}

%%%%%%%%%%%%%%%%%%%%%%%%%%
\subsection{Single-process limits}

We first discuss the limits for each individual process. We show two-dimensional limits, setting the Wilson coefficients not shown to zero, as well as one-dimensional profiled limits. The corresponding one-dimensional constraints, where we set all other Wilson coefficients to zero, are collected in App.~\ref{app:additional_results}.

%%%%
\subsubsection*{\texorpdfstring{$WZ$}{WZ} production}

\begin{figure}[t]
    \centering
    \includegraphics[width=.7\linewidth]{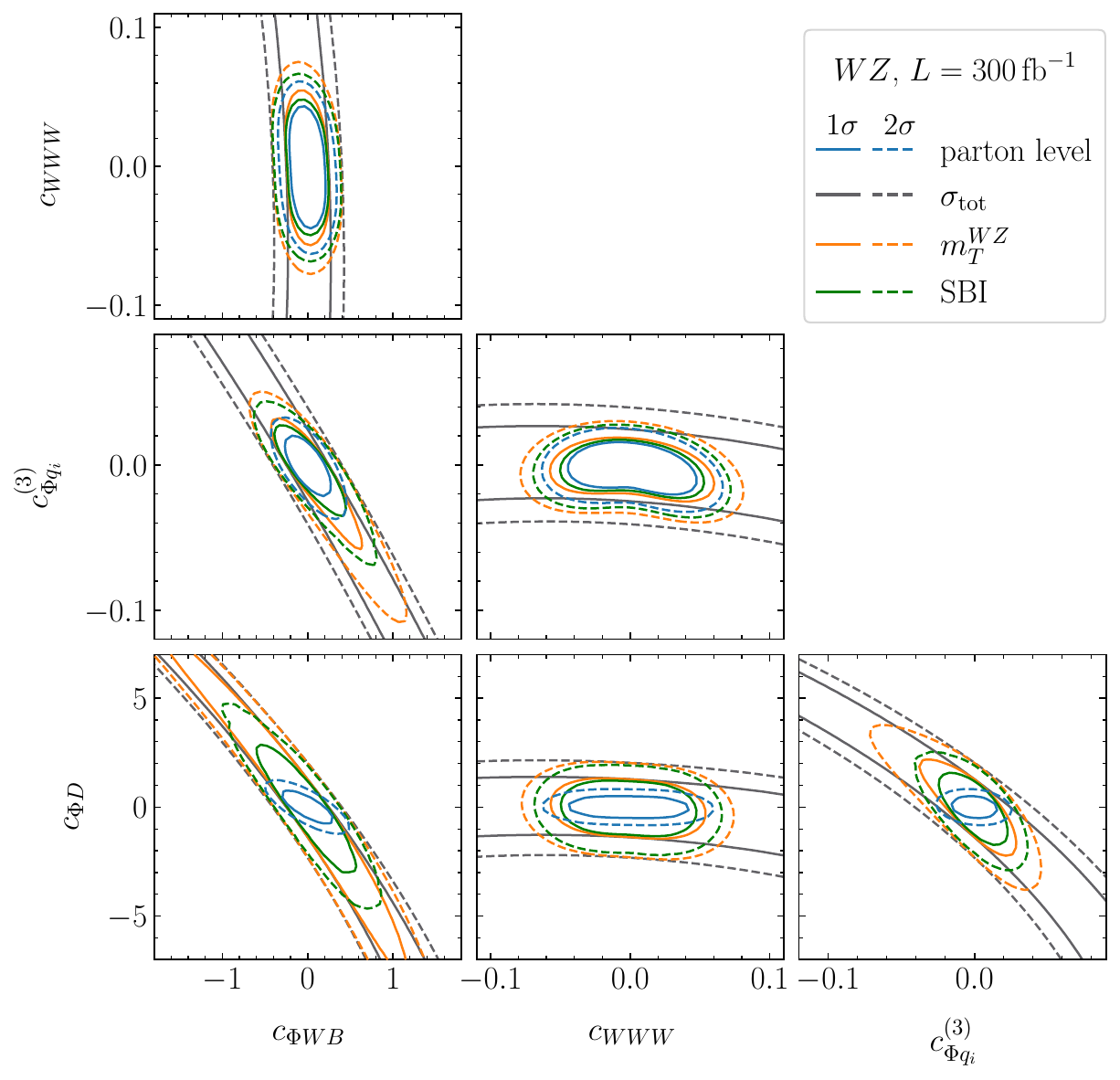}
    \caption{Expected two-dimensional constraints from \textbf{$WZ$ production} setting the not-shown Wilson coefficient to zero.}
    \label{fig:WZ_triangular}
\end{figure}

We start with the $WZ$ process for which we consider the effect of four SMEFT operators. The associated two-dimensional limits, setting the third and fourth Wilson coefficient to zero, are shown in Fig.~\ref{fig:WZ_triangular}. For each result, solid and dashed contours indicate the one-sigma and two-sigma regions, respectively.

%----------------------------
%----------------------------

%----------------------------
\begin{figure}[b!]
    \centering
    \includegraphics[width=0.7\linewidth]{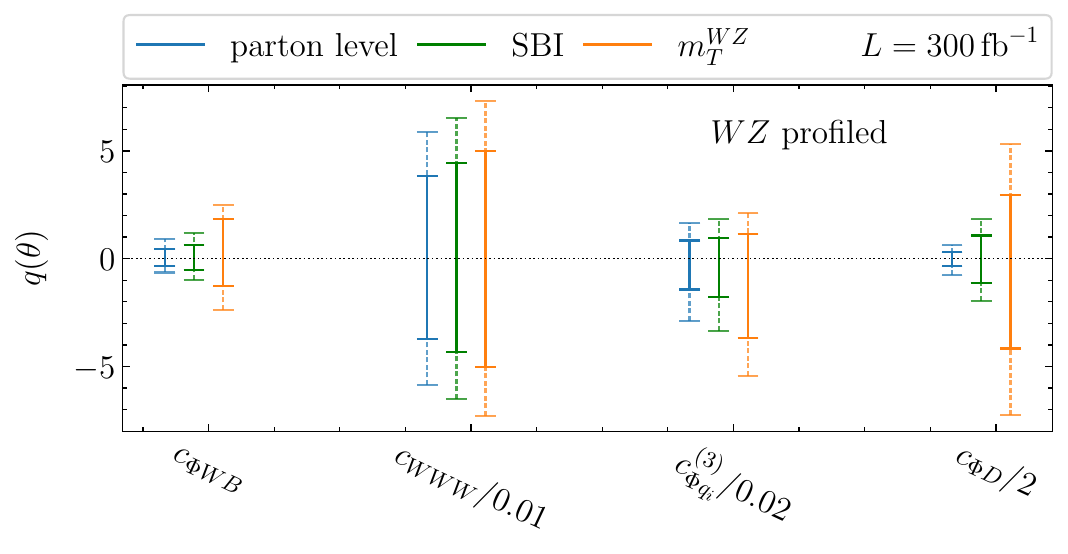}
    \caption{Expected profiled one-parameter confidence intervals for $WZ$ production. The small horizontal lines indicate the one- and two-sigma confidence intervals.}
    \label{fig:WZ_limits}
\end{figure}
%----------------------------

The results agree with Ref.~\cite{Bahl:2024meb}, modulo the additional \OHD operator. The total rate limits shown as grey lines are least constraining, featuring flat directions for all 2D combinations. These flat directions are mostly lifted by kinematic information in the histogram approach (orange dashed). In the \cHWB--\cHD plane, the histogram approach still struggles to disentangle the two operators. Using SBI tightens the limits in comparison to the histogram approach, with the largest improvements in the \cHWB--\cHD, \cHWB--\cHq, and \cHq--\cHD planes. The SBI limits are, as expected, weaker than the parton-level limits (blue lines), which we show as a check of the SBI results.

Moreover, in Fig.~\ref{fig:WZ_limits} we show the single parameter constraints profiled over all Wilson coefficients not shown. We compare parton-level constraints (blue), SBI constraints (green), and histogram-based constraints (orange). We do not show the rate-only results, where the profiling wipes out all constraints. Again, the SBI result is, as expected, weaker than the parton-level constraint. It, however, consistently outperforms the histogram-based limits in particular for \cHWB and \cHD. Looking at Fig.~\ref{fig:WZ_triangular}, we can trace this back to the partial degeneracy between \cHWB and \cHD. The SBI approach breaks this degeneracy more effectively than the histogram-based approach, since it exploits the full dimensionality of the phase space and not only one lower-dimensional summary statistic.

%%%%
\subsubsection*{\texorpdfstring{$WW$}{WW} production}

%----------------------------
\begin{figure}[t]
    \centering
    \includegraphics[width=.7\linewidth]{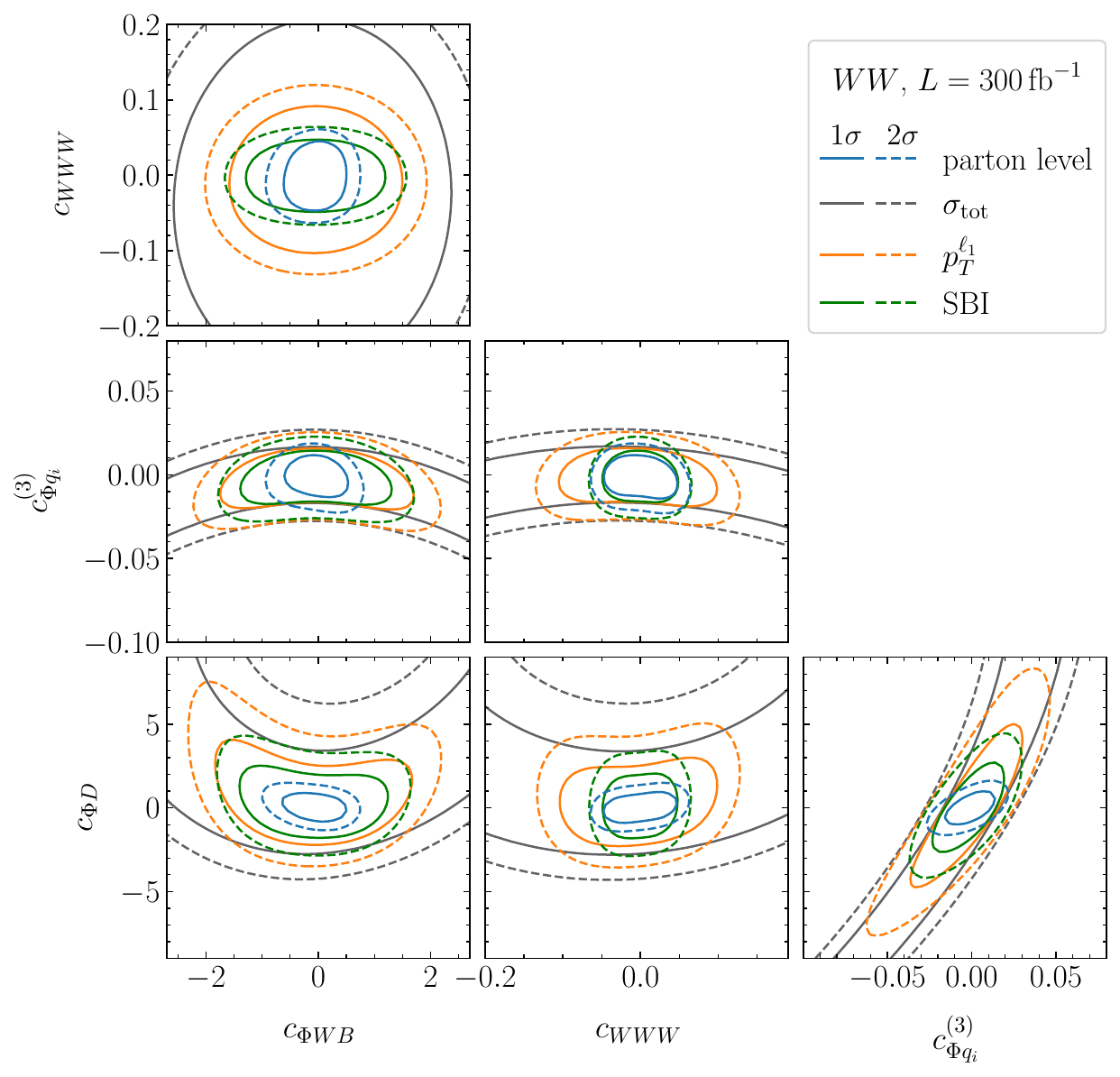}
    \caption{Expected two-dimensional constraints from \textbf{$WW$ production} setting the not-varied Wilson coefficients to zero.}
    \label{fig:WW_triangular}
\end{figure}
%----------------------------

Next, we discuss the $WW$ process, for which we also consider four SMEFT operators. The resulting two-dimensional limits, setting the remaining Wilson coefficients to zero, are shown in Fig.~\ref{fig:WW_triangular}. As for $WZ$, including kinematic information via histograms or SBI drastically improves the constraints, in particular in the \cWWW and \cHWB directions. Moreover, SBI outperforms the histogram approach with the largest improvements being present in the \cHWB--\cHD, the \cHWB--\cWWW, and the \cHq--\cHD planes. Here, SBI again profits from exploiting the full phase space and not only one particular low-dimensional summary statistic.

%----------------------------
\begin{figure}[b!]
    \centering
    \includegraphics[width=0.7\linewidth]{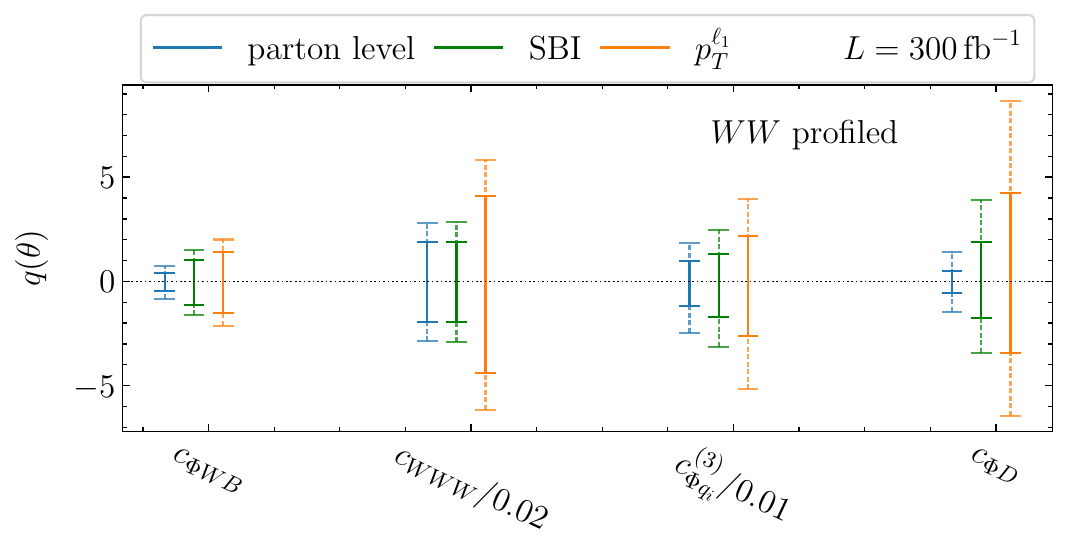}
    \caption{Expected profiled one-parameter confidence intervals for $WW$ production. The small horizontal lines indicate the one- and two-sigma confidence intervals.}
    \label{fig:WW_limits}
\end{figure}
%----------------------------

This is also reflected in the one-dimensional limits profiled over the full three-dimensional parameter space shown in Fig.~\ref{fig:WW_limits}. Using SBI leads to significantly stronger results for all four Wilson coefficients.

%%%%
\subsubsection*{\texorpdfstring{$WH$}{WH} production}

%----------------------------
\begin{figure}[t]
    \centering
    \includegraphics[width=0.5\linewidth]{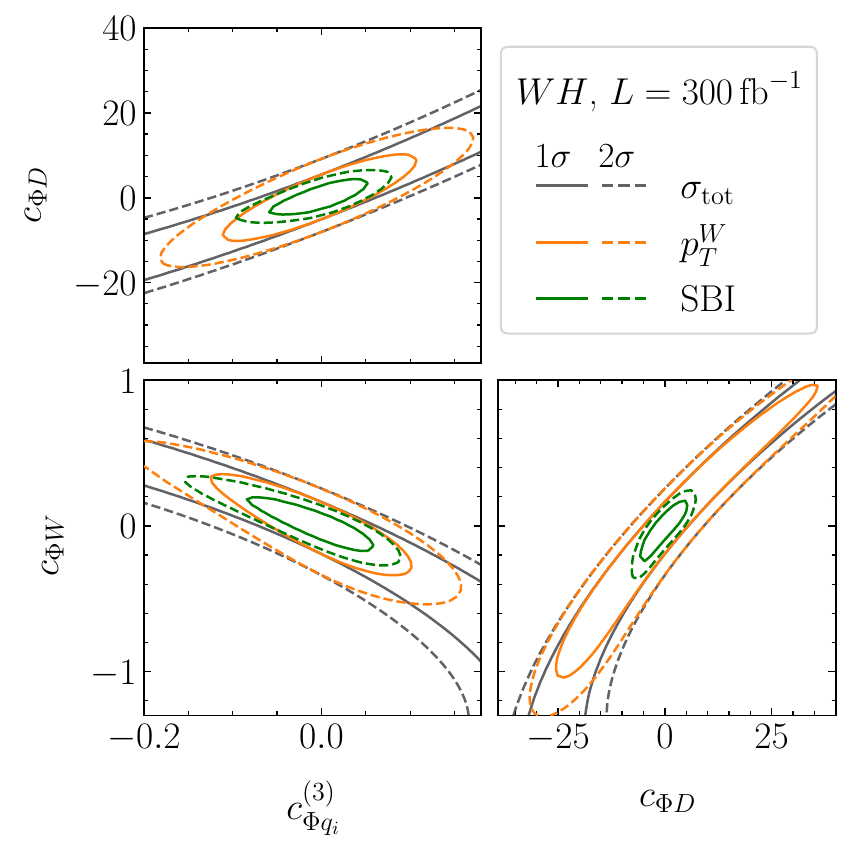}
    \caption{Expected two-dimensional constraints from \textbf{$WH$ production} setting the not-varied Wilson coefficient to zero.}
    \label{fig:WH_triangular_with_bkg}
\end{figure}
%----------------------------

Now, we turn to the first Higgs-production process. $WH$ production is only affected by three of our selected SMEFT operators. The corresponding two-dimensional limits, setting the third relevant Wilson coefficient to zero,  are shown in Fig.~\ref{fig:WH_triangular_with_bkg}. Here, we do not show parton-level  bounds due to the presence of background and the corresponding need to construct the signal--background likelihood ratio. We again observe that the SBI constraints are stronger than the histogram limits. This is particularly evident for the \cHD and \cHW directions in which the $p_T^W$ histogram is not able to extract significant information beyond the total rate. As studied in Ref.~\cite{Brehmer:2019gmn}, also a finer binning in $p_T^W$ or including the total transverse mass as a second histogram variable only slighly improves the sensitivity of the histogram approach.

%----------------------------
\begin{figure}[b!]
    \centering
    \includegraphics[width=0.65\linewidth]{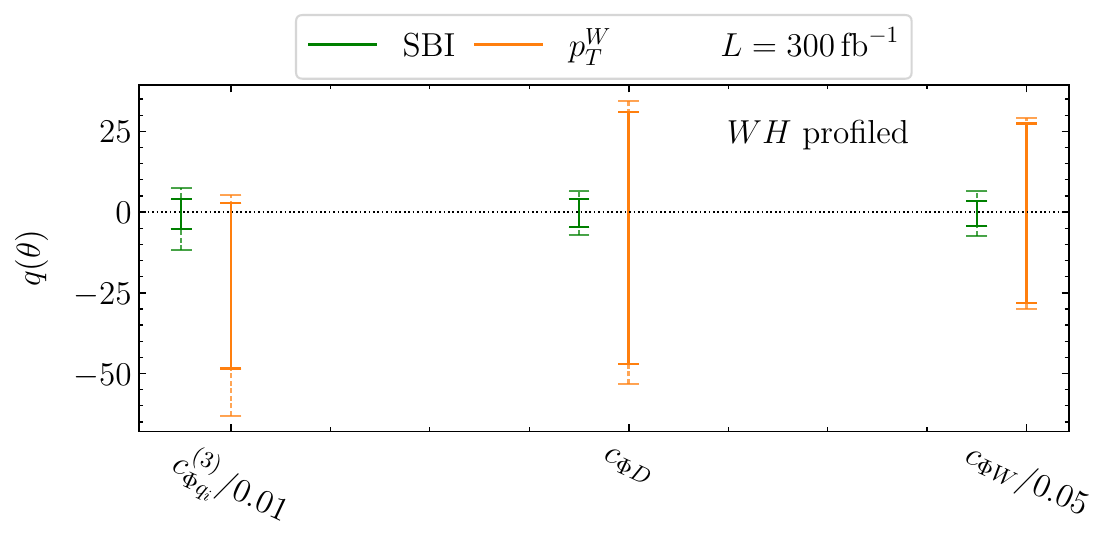}
    \caption{Expected profiled one-parameter confidence intervals for $WH$ production, including backgrounds. The small horizontal lines indicate the one- and two-sigma confidence intervals.}
    \label{fig:WH_limits}
\end{figure}
%----------------------------

Looking at the one-dimensional limits when profiling over the other parameters, as shown in Fig.~\ref{fig:WH_limits}, we also clearly see the advantage of the SBI approach. The additional kinematic information extracted using the SBI approach significantly improves the limit on the three relevant Wilson coefficients. Our findings are in good agreement with Ref.~\cite{Brehmer:2019gmn}. In there, an alternative binning scheme is proposed to bring the histogram results closer to the SBI results.

%%%%
\subsubsection*{\texorpdfstring{$ZH$}{ZH} production}

%----------------------------
\begin{figure}[t]
    \centering
    \includegraphics[width=.8\linewidth]{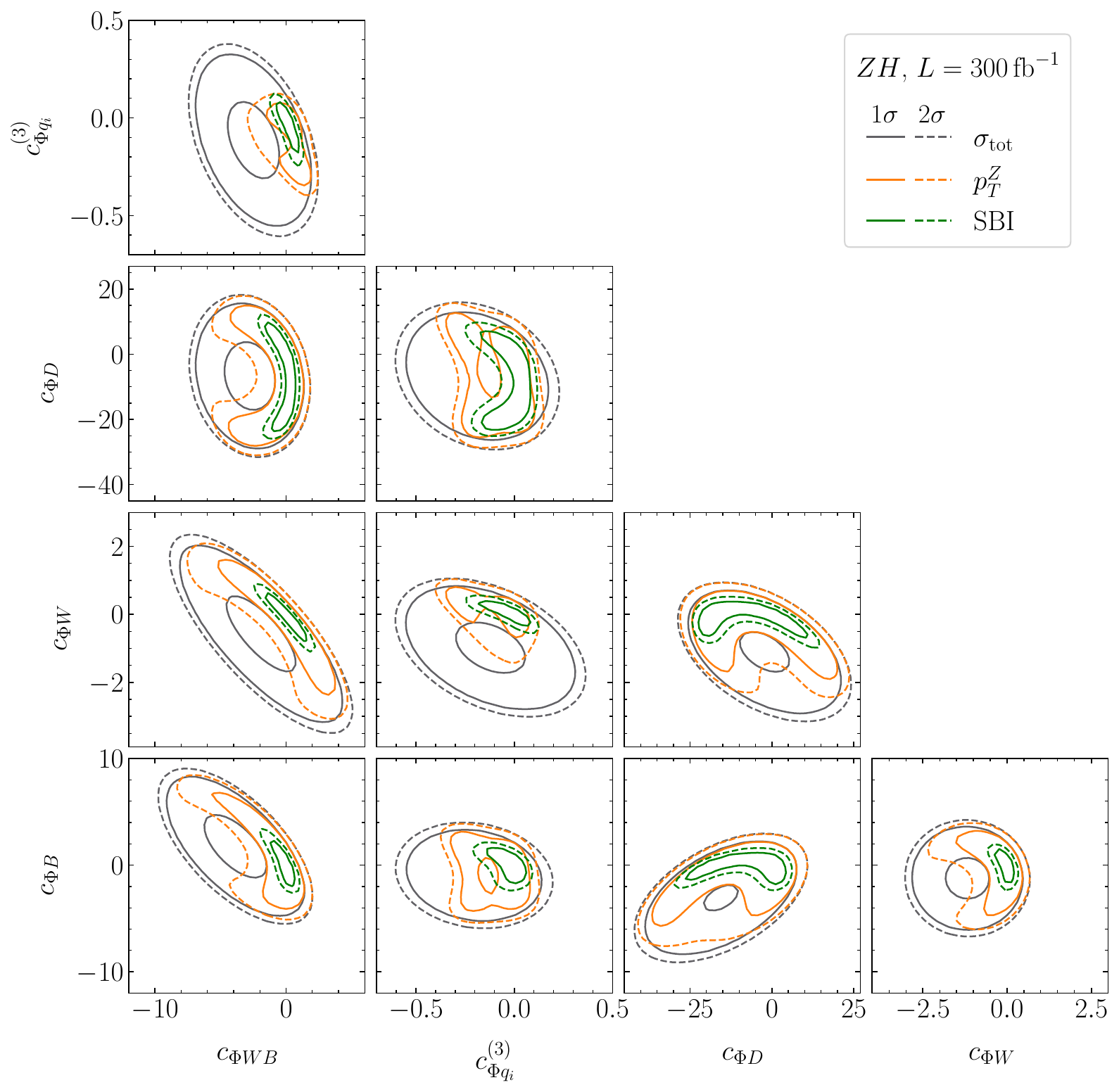}
    \caption{Expected two-dimensional constraints from \textbf{$ZH$ production} setting the not-varied Wilson coefficient to zero. The SM point does not lie in the center of the  parameter planes. For some of the shown constraints, in particular the total rates, the center of the parameter planes are expected to be excluded at the $1\,\sigma$ level resulting in the small central ellipsis. In other words, if the the panel has two solid contours with the same color, the region in between them is allowed at the $1\,\sigma$ level.}
    \label{fig:ZH_triangular}
\end{figure}
%----------------------------

As the final process, we investigate $ZH$ production, which is affected by five of the considered SMEFT operators. The 2D limits, setting the Wilson coefficients not shown to zero, are shown in Fig.~\ref{fig:ZH_triangular}.

%----------------------------
\begin{figure}[b!]
    \centering
    \includegraphics[width=0.7\linewidth]{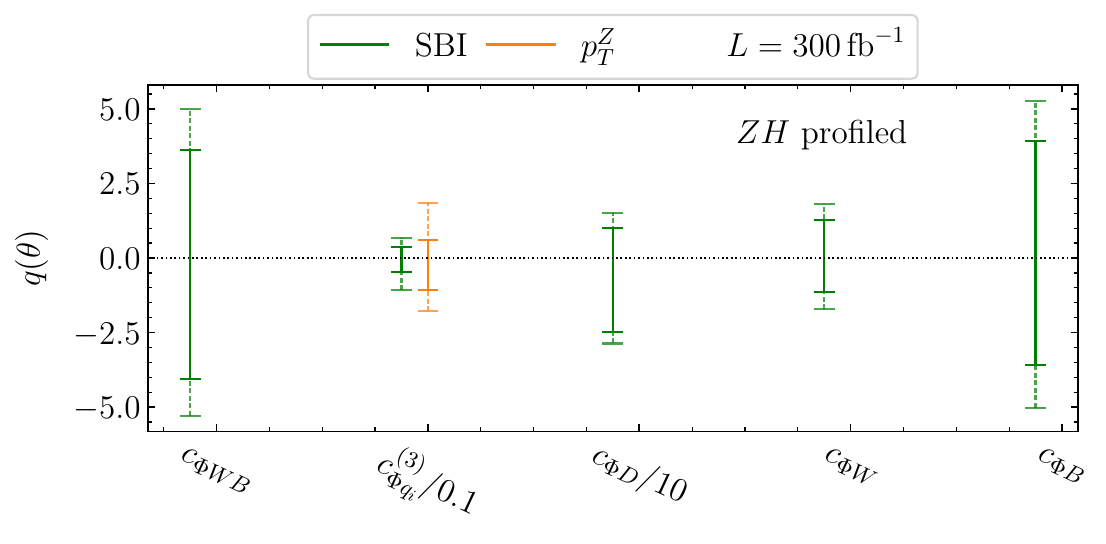}
    \caption{Expected profiled one-parameter confidence intervals for $ZH$ production, including backgrounds. The small horizontal lines indicate the one- and two-sigma confidence intervals. If no confidence interval is shown, the corresponding parameter is not constrained.}
    \label{fig:ZH_limits}
\end{figure}
%----------------------------

First, the total rate constraints are relatively insensitive, making the kinematic information of the histogram and SBI approaches more relevant. This is a consequence of the large background contribution to the signal region. A harder cut on the signal--background classifier score could enhance the rate constraints, but would reduce the kinematic constraints if combined with the kinematic SBI or histogram information. As a consequence of the larger background contributions, the bands observed for the other processes are here enlarged to (partial) ellipses.

For the histogram approach, the constraints get significantly tighter. They become even tighter once the full kinematic information is extracted using the SBI approach. The effect is particularly strong for the \cHW, \cHB, \cHWB, and \cHD directions and the two-dimensional combinations thereof. This improvement is due to the SBI approach being sensitive to the $Z$ boson polarization, disentangling the left-handed and right-handed $Z$ coupling to quarks, see App.~\ref{app:SMEFT_Feynman_rules}. In contrast, the $p_T^Z$ histogram is not sensitive to the $Z$ boson polarization.

In the upper half of the \cHq--\cHD plane, we observe for $\cHD\simeq 5$ and $\cHq \simeq - 0.2$ a small region in which the histogram-based approach is slightly stronger than the SBI limits. This is likely due to imperfect training of the $R_{\cHq\cHD}$ differential cross-section ratio.

The advantages of the SBI approach are even more visible in the one-dimensional profiled limits in Fig.~\ref{fig:ZH_limits}. Since the $p_T^Z$ histogram is not sensitive to the $Z$ polarization, the \cHWB, \cHD, \cHW, and \cHB Wilson coefficients form a degenerate set resulting in no one-dimensional profiled constraint for any of these coefficients. The SBI approach breaks this degeneracy and is able to constrain each of these coefficients even if profiling over the others. Only for \cHq, the histogram-based approach yields a profiled one-dimensional limit, whose lower bound is weaker than the corresponding SBI constraint.

%%%%%%%%%%%%%%%%%%%%%%%%%%
\subsection{Combined limits}

Until now, we have shown that SBI leads to significantly stronger constraints on the Wilson coefficients for each of the considered processes in comparison to the histogram approach. The following results show how these advantages persist when combining all four processes.

As mentioned above, we neglect systematic uncertainties. Moreover, the individual signal and background processes do not overlap within the regions defined by the pre-selection cuts introduced in Sec.~\ref{sec:di-boson_procs}. Therefore, we assume the different processes to be uncorrelated. The combined test statistic is then
\begin{align}
    q_\text{comb}(\theta) = q_{WW}(\theta) + q_{WZ}(\theta) + q_{WH}(\theta) + q_{ZH}(\theta) \;.
\end{align}
Based on this combined test statistic, we derive the combined confidence regions. We note that this simplification will in general not hold at the LHC.

%----------------------------
\begin{figure}[t]
    \centering
    \includegraphics[width=.95\linewidth]{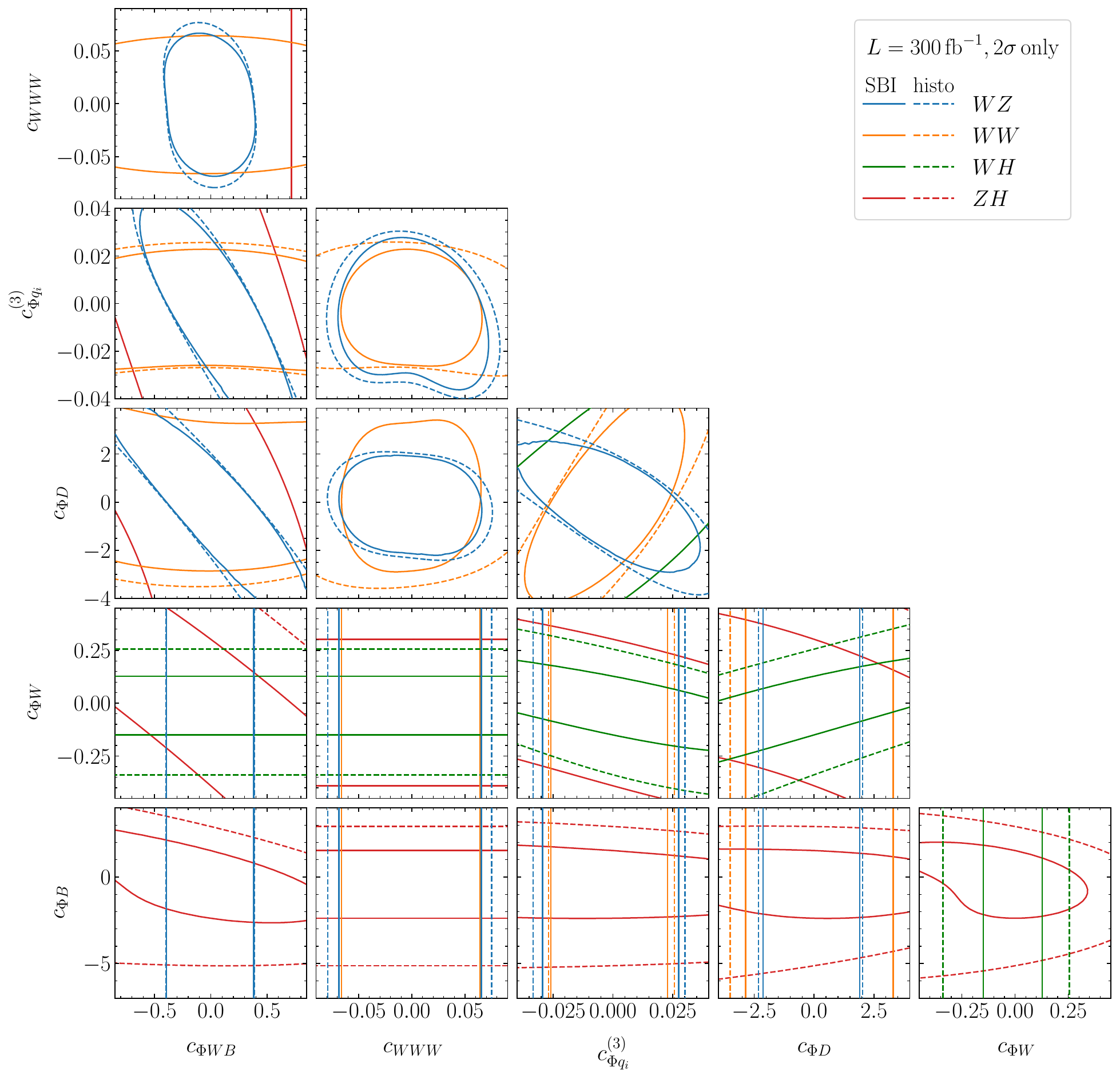}
    \caption{Expected two-dimensional 95\% C.L.\ constraints setting the not-varied Wilson coefficient to zero. We compare the limits of the four single processes using either the histogram approach (dashed lines) or SBI (solid lines).}
    \label{fig:combined_limits_processes}
\end{figure}
%----------------------------

Before showing the combined limits, we investigate in Fig.~\ref{fig:combined_limits_processes} the contribution of each process. Here, we focus only on the SBI approach (solid lines) and the histogram approach (dashed lines). The different colors indicate the processes. Each process is important to constrain at least one of the directions. For example, the \cHWB direction is mostly constrained by $WZ$ production; the \cHW direction, mostly by $WH$ production; the \cHB direction, mostly by $ZH$ production; and, the \cHq direction, mostly by $WW$ production. This clearly shows the benefit of a combined analysis of all four di-boson processes. We also observe that SBI is more effective than the histogram approach in constraining most of the directions. This then translates into more stringent combined constraints.

%----------------------------
\begin{figure}[t]
    \centering
    \includegraphics[width=.95\linewidth]{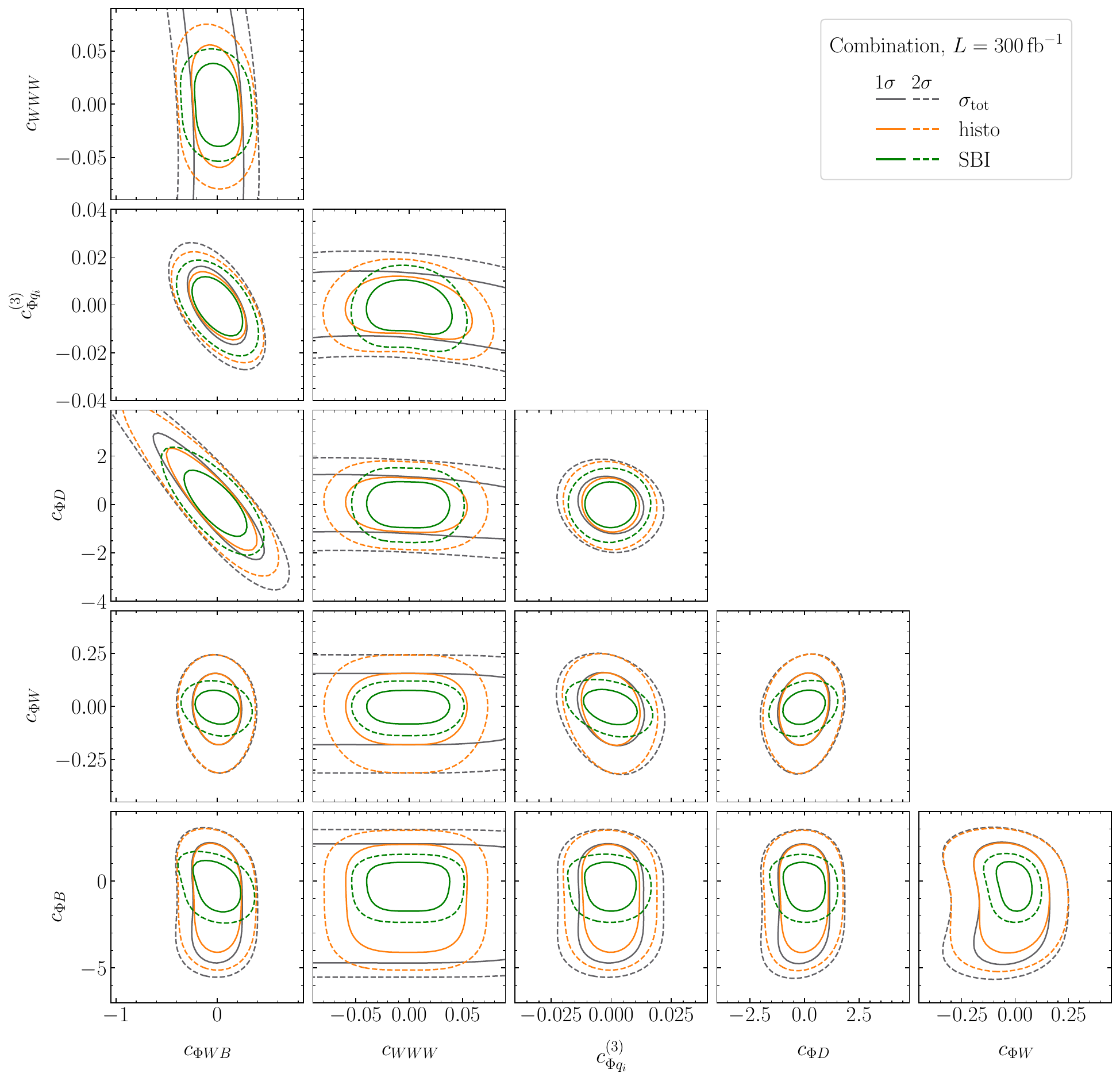}
    \caption{Expected two-dimensional constraints from the combined di-boson processes setting the not-varied Wilson coefficient to zero. We compare results based on SBI, histogram-based inference, and only the total rate information.}
    \label{fig:combined_triangular}
\end{figure}
%----------------------------

We show the combined 2D limits, setting all other Wilson coefficients to zero, in Fig.~\ref{fig:combined_triangular}. As before, the SBI approach consistently outperforms the histogram-based limits. In some parameter planes, e.g. the \cHWB--\cHq plane, the differences are relatively small, because these directions are well constrained by the histograms for at least one of the considered processes. In other parameter planes, e.g. the \cWWW--\cHB plane, we find larger improvements with the SBI approach clearly outperforming the histogram limits. This is in particular true if the histogram approach is not able to extract information beyond the total rate, as seen e.g.\ in the \cHW, \cHB, and \cHD directions. One example is the \cHWB--\cHW plane for which the additional information from $WH$ and $ZH$ in the SBI approach allows to significantly improve the constraints in comparison to the histogram approach.

%----------------------------
\begin{figure}[t]
    \centering
    \includegraphics[width=.8\linewidth]{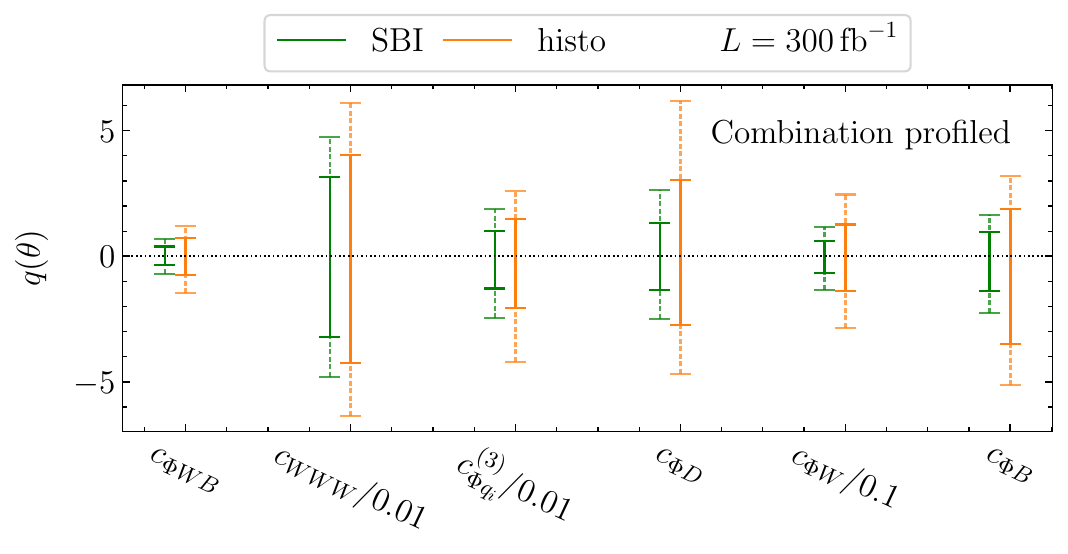}
    \caption{Expected profiled one-parameter confidence intervals combining all four di-boson processes, including backgrounds. The small horizontal lines indicate the one- and two-sigma confidence intervals.}
    \label{fig:combined_limits}
\end{figure}
%----------------------------

Finally, in Fig.~\ref{fig:combined_limits} we show the one-dimensional limits profiling over all other Wilson coefficients using the combination of all four considered processes. For all six considered Wilson coefficients, we see significant improvements of the SBI limits over the histogram-based limits. While the combination of the different processes avoids flat directions in the histogram limits, as observed for $WH$ production in Fig.~\ref{fig:WH_limits}, combining different channels does not bring the sensitivity of the histogram approach to the same level as the SBI approach. Differences are particularly striking for \cHD, \cHW, \cHWB, and \cHB. As discussed for $ZH$ production, these Wilson coefficients are strongly correlated if the analysis is not sensitive to the polarization of the $Z$ boson. The SBI limits on these coefficients are a factor of $\sim 2$ stronger than the histogram-based limits. To reach the same sensitivity in the histogram approach, a factor $\sim 4$ more data would be needed if we assume a naive scaling of the limits with the inverse square root of the luminosity. But even for the \cWWW and \cHq Wilson coefficients, to which the chosen histogram observables are more directly sensitive, the SBI approach is $\sim 30\%$ more sensitive, corresponding to a factor of $\sim 1.8$ more data.

%%%%%%%%%%%%%%%%%%%%%%%%%%%%%%%%%%%%%%%%%%%%%%%%%%%%%%
\section{Conclusions}
\label{sec:conclusions}

Neural simulation-based inference (SBI) is a key methodology if we want to unleash the full potential of the current and future LHC runs. In comparison to traditional approaches like histogramming, it does not rely on binned low-dimensional summary statistics. Exploiting the full high-dimensional event information not only leads to an improved sensitivity to single theory parameters but is also more effective in disentangling different parameters.

This is particularly evident in the SMEFT context. Often, a single process is affected by many different operators, each of which changes the kinematic distributions in a slightly different manner. We have shown in this paper that SBI clearly outperforms histogram-based methods in such scenarios, focusing on four different di-boson production channels, $WW$, $WZ$, $WH$, and $ZH$ production, and six SMEFT operators. For each process, the SBI limits are significantly tighter than the histogram-based limits. In particular, SBI excels at lifting degeneracies between two or more Wilson coefficients, resulting in significantly stronger one-dimensional profiled limits. We also checked the coverage of the learned likelihoods finding good agreement with the nominal confidence level. Our analysis is simplified since we neglect systematic uncertainties as well as subleading backgrounds and use a fast detector simulation. We, however, expect SBI to handle a more realistic setting better than the histogram-based approach.

Apart from showing that SBI outperforms the histogram approach for single-process limits, which has been shown before, we took a further step to demonstrate the potential of SBI. In particular, it has so far been unclear if the superior sensitivity of SBI persists when multiple processes are combined in a global analysis, as it is common practice for constraining SMEFT parameters. To answer this question, we have combined the limits of the four di-boson processes using a simplified procedure. In this more global analysis context, the SBI shows persistent advantages over the  histogram-based limits. It is significantly more sensitive, even for Wilson coefficients which strongly alter the distributions of the chosen histogram observables. Assuming a naive scaling of the constraints with the inverse square root of the luminosity, the SBI improvements over histogram-based inference corresponds to a factor two or more in luminosity. As a first step towards an actual global SBI fit, also the histogram-based and SBI-based likelihoods can be combined.

%%%%%%%%%%%%%%%%%%%%%%%%%%%%%%%%%%%%%%%%%%%%%%%%%%%
\section*{Acknowledgments}

This research is supported by the Deutsche Forschungsgemeinschaft (DFG, German Research Foundation) under grant 396021762--TRR~257: \textsl{Particle Physics Phenomenology after the Higgs Discovery}, and through Germany's Excellence Strategy EXC~2181/1 -- 390900948 (the \textsl{Heidelberg STRUCTURES Excellence Cluster}). We would also like to thank the Baden-W\"urttem\-berg Stiftung for financing through the program \textsl{Internationale Spitzenforschung}, project \textsl{Uncertainties – Teaching AI its Limits} (BWST\_ISF2020-010). We acknowledge support by the state of Baden-Württemberg through bwHPC and the German Research Foundation (DFG)
through grant no INST 39/963-1 FUGG (bwForCluster NEMO).

%%%%%%%%%%%%%%%%%%%%%%%%%%%%%%%%%%%%%%%%%%%%%%%%%%
\appendix

\clearpage
%%%%%%%%%%%%%%%%%%%%%%%%%%%%%%%%%%%%%%%%%%%%%%%%%%%%%%
\section{Relevant SMEFT Feynman rules}
\label{app:SMEFT_Feynman_rules}

Following Ref.~\cite{Dedes:2017zog,Degrande:2020evl}, we list here all relevant SMEFT Feynman rules neglecting $\mathcal{CP}$ and flavour violation:
\begin{align}
    c(u^{f_1}u^{f_2}Z_{\mu}) ={}& \frac{2i e s_{W}}{3 c_W}\delta_{f_1f_2}\gamma_\mu P_R + i e \left(\frac{s_W}{6 c_W} - \frac{c_W}{2s_W}\right)\delta_{f_1f_2} \gamma_\mu P_L \notag \\
    & - \frac{i e}{2 c_W s_W}\cHq \frac{v^2}{\Lambda^2}\delta_{f_1f_2}\gamma_\mu P_L  - \frac{2}{3}i e \cHWB \frac{v^2}{\Lambda^2}\delta_{f_1f_2} \gamma_\mu P_L  \notag \\
    & - \frac{1}{6}i e \cHD\frac{v^2}{\Lambda^2}\delta_{f_1f_2}\gamma_\mu\left[\left(\frac{1}{4c_W s_W} + \frac{c_W}{s_W}\right)P_L + \left(\frac{1}{c_W s_W} + \frac{c_W}{s_W}\right)P_R\right]\;,\notag \\
    c(d^{f_1}d^{f_2}Z_{\mu}) ={}& -\frac{i e s_{W}}{3 c_W}\delta_{f_1f_2}\gamma_\mu P_R + i e \left(\frac{s_W}{6 c_W} + \frac{c_W}{2s_W}\right)\delta_{f_1f_2} \gamma_\mu P_L \notag \\
    & + \frac{i e}{2 c_W s_W}\cHq \frac{v^2}{\Lambda^2}\delta_{f_1f_2}\gamma_\mu P_L 
    % + \frac{i e}{2 s_w c_W} \frac{v^2}{\Lambda^2} \cHbox P_L
     \notag \\
    & - \frac{1}{6}i e \cHD\frac{v^2}{\Lambda^2}\delta_{f_1f_2}\gamma_\mu\left[\left(\frac{1}{4c_W s_W} + \frac{3c_W}{2s_W}\right)P_L + \left(\frac{1}{4c_W s_W} + \frac{3c_W}{2s_W}\right)P_R\right]\;,\notag \\
    c(u^{f_1}\bar d^{f_2}W^+_{\mu}) ={}&\frac{i e}{\sqrt{2}s_W}\delta_{f_1f_2}\gamma_{\mu}P_L + \frac{i e}{\sqrt{2}s_W}\cHq\frac{v^2}{\Lambda^2}\delta_{f_1f_2}\gamma_{\mu}P_L\;, \notag \\
    c(u^{f_1}u^{f_2}Z_{\mu} h) ={}& -  \frac{i e}{c_W s_W}  \cHq \frac{v}{\Lambda^2} \delta_{f_1f_2}\gamma_\mu P_L\;, \notag \\
    c(d^{f_1}d^{f_2}Z_{\mu} h) ={}&   \frac{i e}{c_W s_W}  \cHq \frac{v}{\Lambda^2} \delta_{f_1f_2}\gamma_\mu P_L +   \frac{i e}{c_W s_W}  \cHD \frac{v}{\Lambda^2} \delta_{f_1f_2}\gamma_\mu P_R\;, \notag \\
    c(u^{f_1}d^{f_2}W^+_{\mu} h) ={}& \sqrt{2} \frac{i e}{s_W}  \cHq \frac{v}{\Lambda^2} \delta_{f_1f_2}\gamma_\mu P_L\;, \notag \\
    c(hW^+_{\mu}W^-_{\nu}) ={}& \frac{i e^2}{2 s_W^2} v g_{\mu\nu}\left(1 
    %+ \cHbox \frac{v^2}{\Lambda^2} 
    - \frac{1}{4}\cHD \frac{v^2}{\Lambda^2}\right)\notag \\
    &+4 i \cHW \frac{v}{\Lambda^2}(p_{2\mu} p_{3\nu} - p_2\cdot p_3 g_{\mu\nu}) \;,  \notag \\
    c(hZ^+_{\mu}Z^-_{\nu}) ={}& \frac{i e^2 v}{2 c_W^2 s_W^2} g_{\mu\nu}\left(1 
    %+ \cHbox\frac{v^2}{\Lambda^2} 
    + \frac{1}{4}\cHD\frac{v^2}{\Lambda^2}\right) \notag \\
    &+ 4 i \left(c_W^2 \cHW + c_W s_W \cHWB + s_W^2 \cHB\right)\frac{v}{\Lambda^2}\left(p_{2\mu} p_{3\nu} - p_2\cdot p_3 g_{\mu\nu}\right)  \;,   \notag \\
    c(h\gamma_{\mu}Z_{\nu}) ={}&  2 i \left(2 c_W s_W \cHW + (s_W^2 - c_W^2) \cHWB + 2 c_W s_W \cHB\right)\notag\notag \\
    &\cdot\frac{v}{\Lambda^2}\left(p_{2\mu} p_{3\nu} - p_2\cdot p_3 g_{\mu\nu}\right) \;, \notag \\
    c(W^-_{\mu_1} W^+_{\mu_2} Z_{\mu_3}) =&{} - i \frac{e c_W}{s_W} \left(1 + \frac{1}{4}\frac{v^2}{\Lambda^2}\cHD\right)\left(g_{\mu_1\mu_2}(p_1 - p_3)_{\mu_3} + g_{\mu_2\mu_3}(p_2 - p_3)_{\mu_1} \right.\notag \\
    &\left. \hspace{4cm}+ g_{\mu_1\mu_3}(p_3 - p_1)_{\mu_2}\right) \notag \\
    & - 6 i c_W \frac{\cWWW}{\Lambda^2}\left(p_{3,\mu_1}p_{1,\mu_2}p_{2,\mu_3} - p_{2,\mu_1}p_{3,\mu_2}p_{1,\mu_3} \right.\notag \\
    &\left.\hspace{2.5cm}+ g_{\mu_1\mu_2}(p_{1,\mu_3}p_2\cdot p_3 - p_{2,\mu_3} p_1\cdot p_3)  \right.\notag \\
    &\left.\hspace{2.5cm} + g_{\mu_2\mu_3}(p_{2,\mu_1} p_1\cdot p_3 - p_{3,\mu_1}p_1\cdot p_2)   \right.\notag \\
    &\left.\hspace{2.5cm}+ g_{\mu_1\mu_3}(p_{3,\mu_2} p_1\cdot p_2 - p_{1,\mu_2}p_2\cdot p_3)\right) \notag \\
    & - i e \cHWB \frac{v^2}{\Lambda^2}\left(g_{\mu_1\mu_2}(p_1 - p_2)_{\mu_3} - g_{\mu_1\mu_3}p_{1,\mu_2} + g_{\mu_2\mu_3}p_{2,\mu_1}\right)\;.
\end{align}
As electroweak input parameters, we use $M_W$, $M_Z$, and $G_F$. The parameters $v$, $e$, $c_W$, and $s_W$ are then fixed via
\begin{align}
    v &= \frac{1}{\sqrt{\sqrt{2}G_F}}\;,\quad e = \frac{2M_W s_{W}}{v}\,,\quad c_W = \frac{M_W}{M_Z}\;,\quad s_{W} = \sqrt{1 - c_W^2}\;.
\end{align}
%

%%%%%%%%%%%%%%%%%%%%%%%%%%%%%%%%%%%%%%%%%%%%%%%%%%%%%%
\section{Likelihood coverages}
\label{app:coverage}

%----------------------------
\begin{figure}[b!]
    \centering
    \includegraphics[width=1\linewidth]{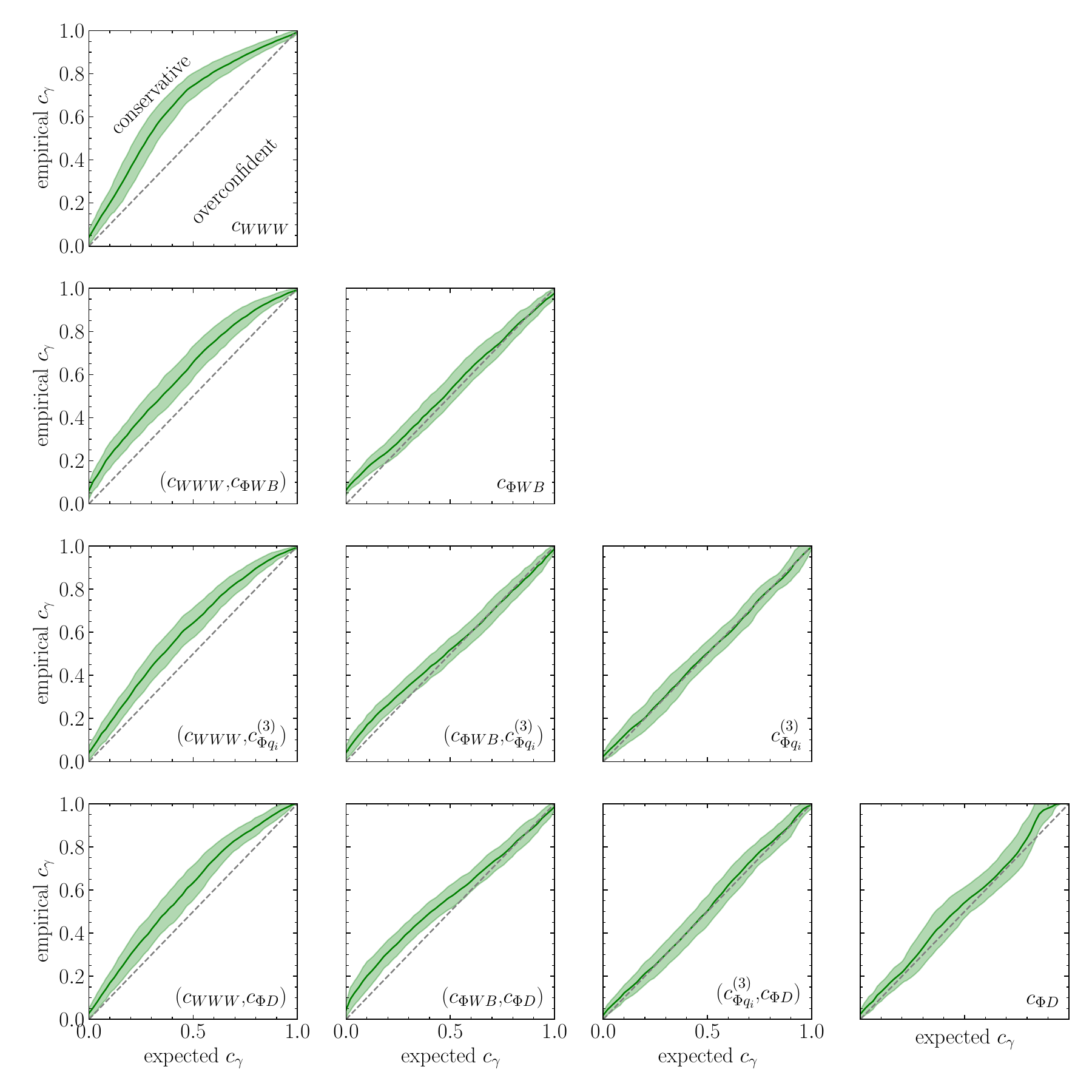}
    \caption{Two-dimensional coverage curves for all pairs of Wilson coefficients for the $WZ$ dataset.}
    \label{fig:WZ_coverage}
\end{figure}
%----------------------------

For computing the coverage of the learned likelihoods, we restrict us for computational efficiency to two-dimensional parameter scans. Since the likelihood is a second-order polynomial in the Wilson coefficients, two-dimensional parameter planes are sufficient to test all directions and correlations of the overall likelihood. Analogously, testing the coverage of the single-process likelihoods is also sufficient to test the coverage of the combined likelihood.

\begin{figure}[t]
    \centering
    \includegraphics[width=1\linewidth]{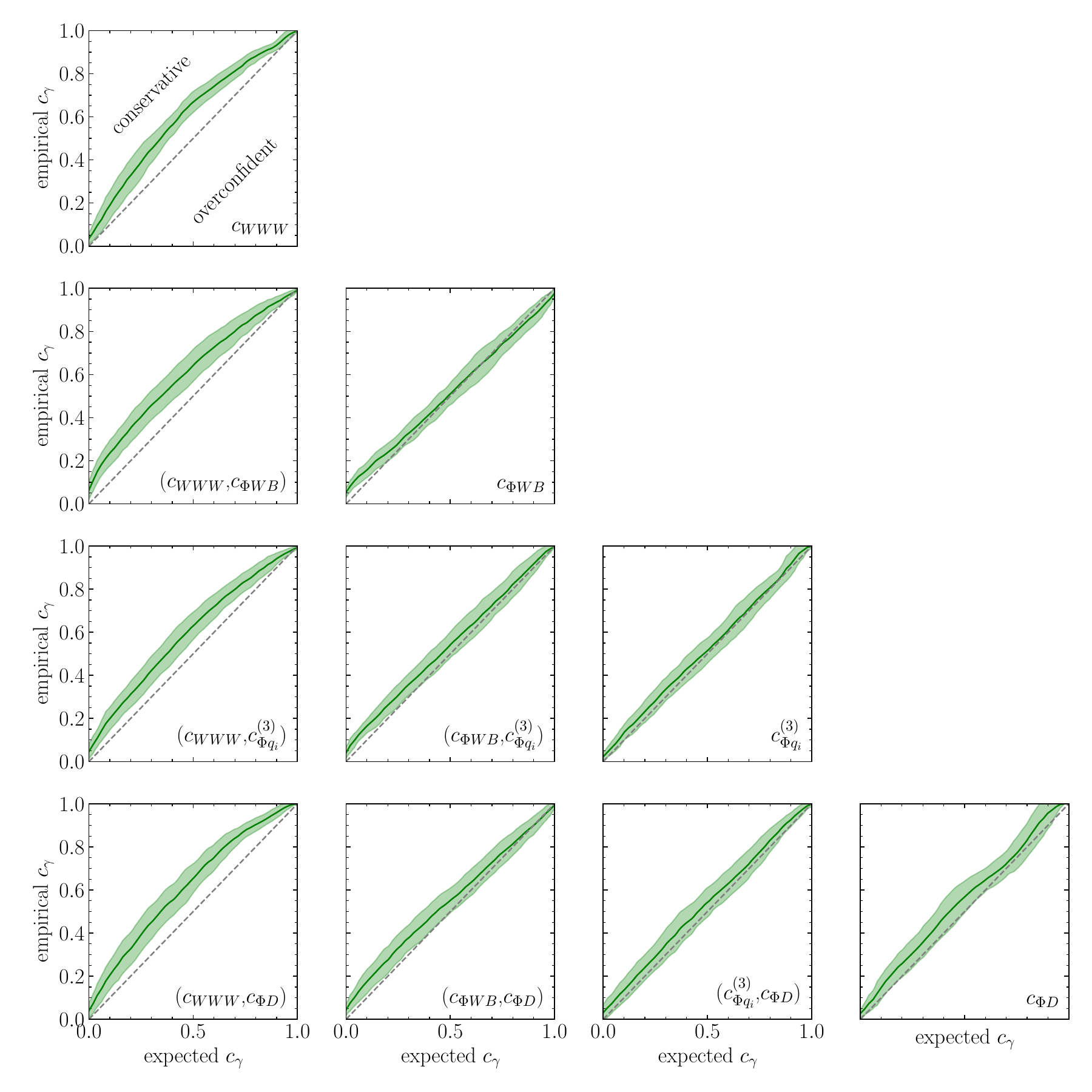}
    \caption{Two-dimensional coverage curves for all pairs of Wilson coefficients for the $WW$ dataset.}
    \label{fig:WW_coverage}
\end{figure}

\begin{figure}[t]
    \centering
    \includegraphics[width=0.6\linewidth]{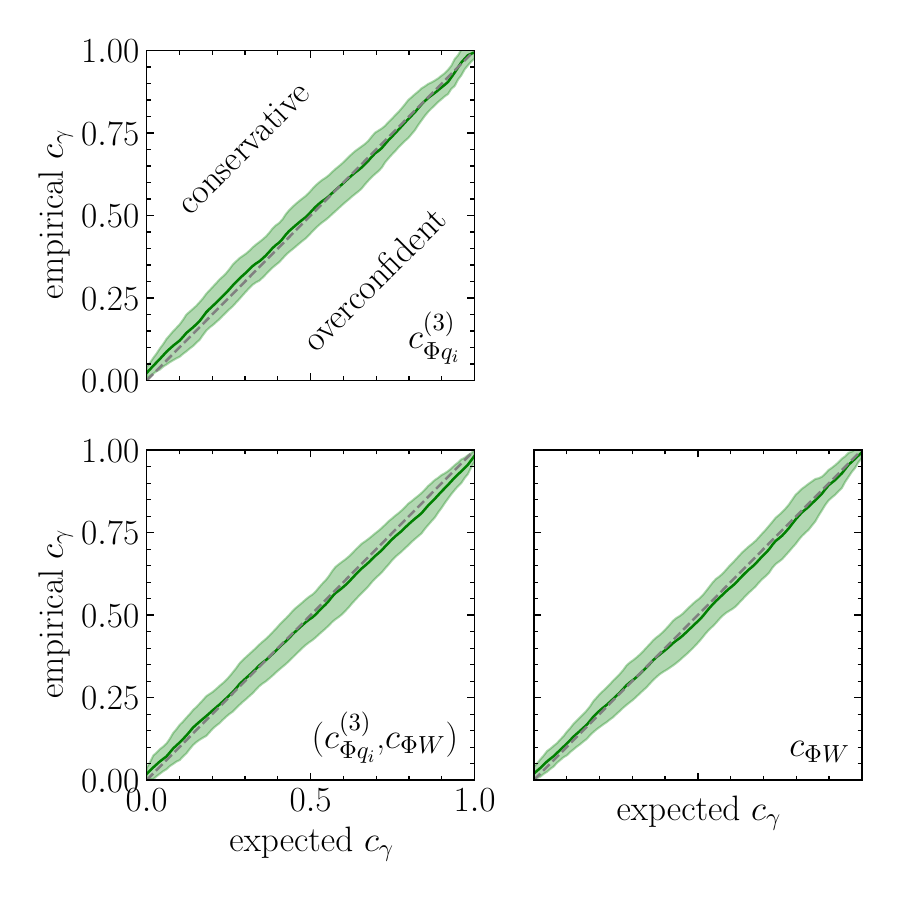}
    \caption{Two-dimensional coverage curves for all pairs of Wilson coefficients for the $WH$ dataset. Here, we do not show parameter combinations involving \cHD, since \cHD is only constrained by total rate information.}
    \label{fig:WH_coverage}
\end{figure}

\begin{figure}[t]
    \centering
    \includegraphics[width=1\linewidth]{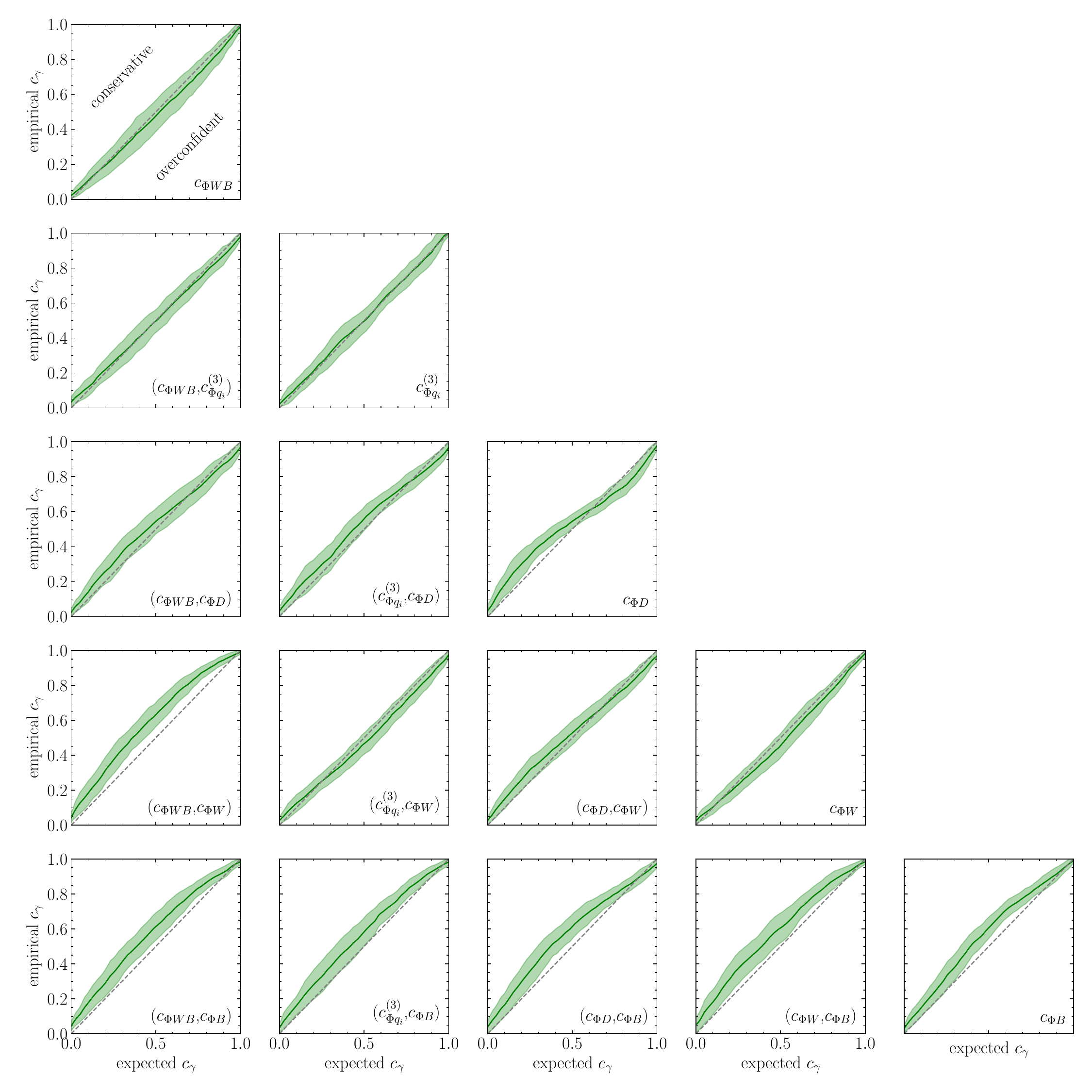}
    \caption{Two-dimensional coverage curves for all pairs of Wilson coefficients for the $ZH$ dataset. }
    \label{fig:ZH_coverage}
\end{figure}

The resulting coverage plots showing the empirical coverage against the expected coverage are shown in Figs.~\ref{fig:WZ_coverage}--\ref{fig:ZH_coverage}. For most of the parameter planes, the coverage curves lie, as expected, on the diagonal. Only for a few directions --- e.g., \cWWW in $WW$ production ---, we observe that the learned likelihood is slightly underconfident resulting in more conservative limits. In these case, an improved extraction of the likelihood --- i.e., more training data --- would help to further tighten the expected constraints.

%----------------------------
%----------------------------

%----------------------------
%----------------------------

%----------------------------
%----------------------------

\clearpage
%%%%%%%%%%%%%%%%%%%%%%%%%%%%%%%%%%%%%%%%%%%%%%%%%%%%%%
\section{Single parameter results}
\label{app:additional_results}

%----------------------------
\begin{figure}[t]
    \centering
    \includegraphics[width=0.5\linewidth]{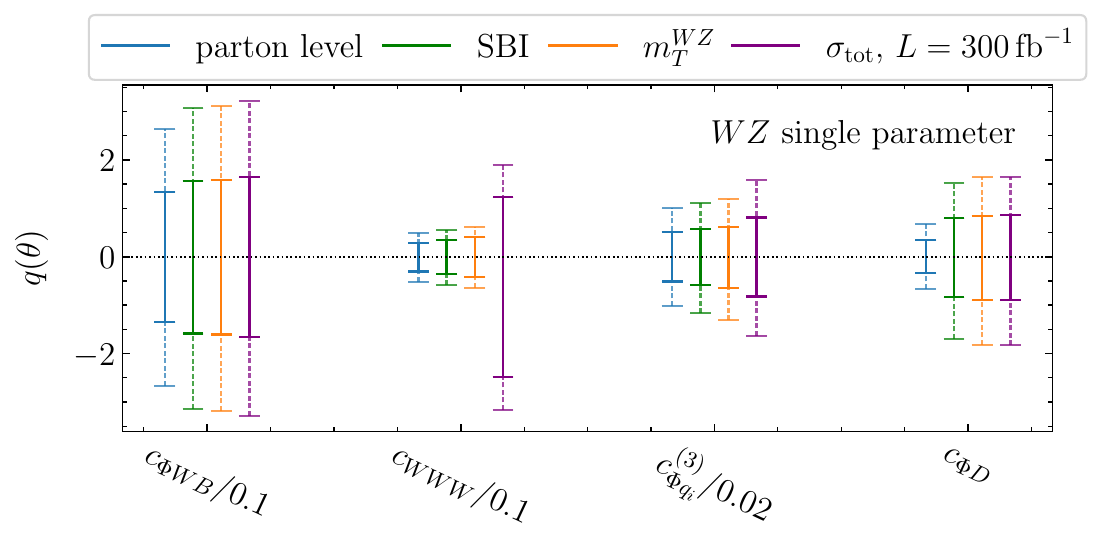}
    \includegraphics[width=0.5\linewidth]{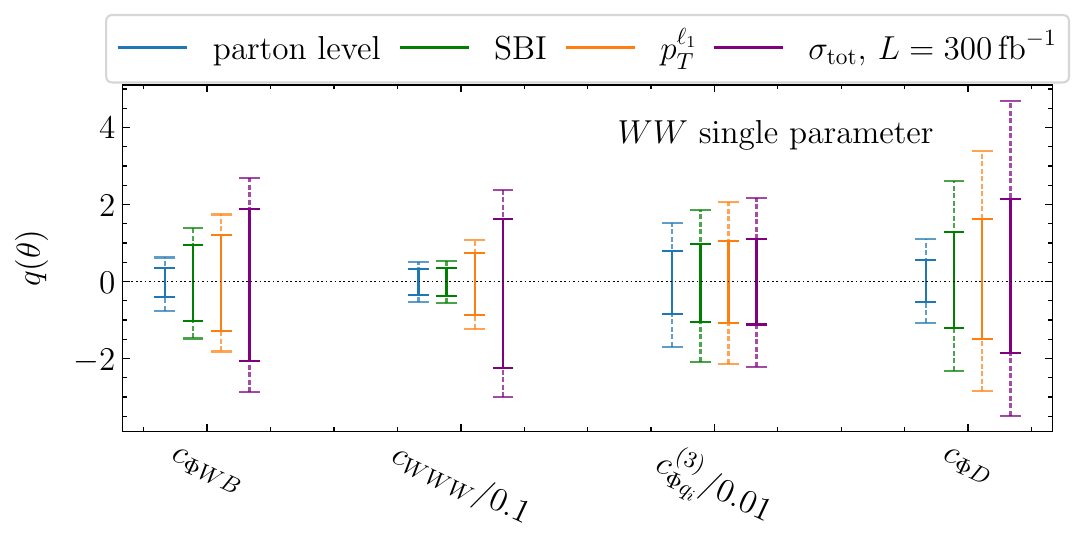}
    \includegraphics[width=0.5\linewidth]{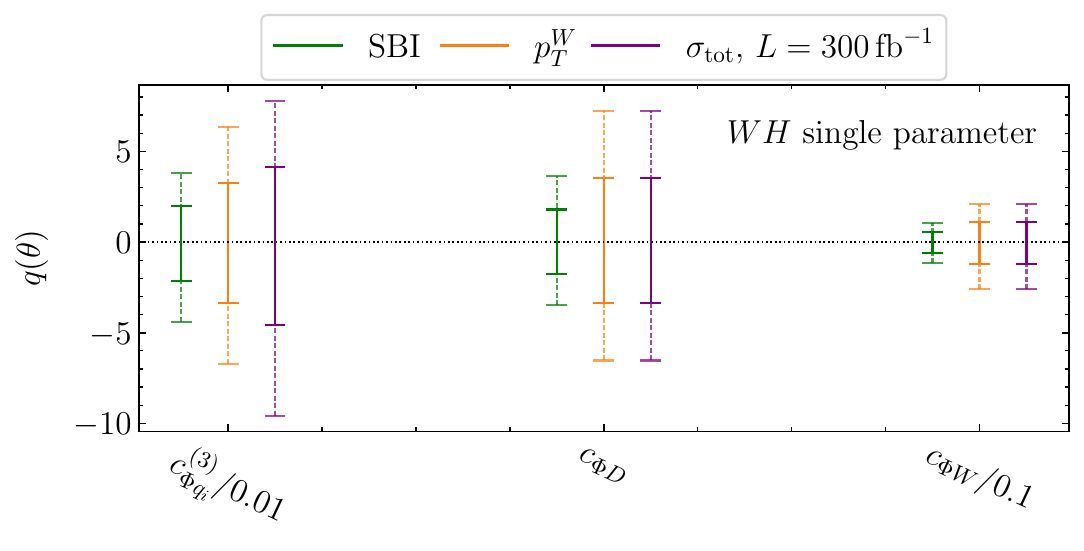}
    \includegraphics[width=0.5\linewidth]{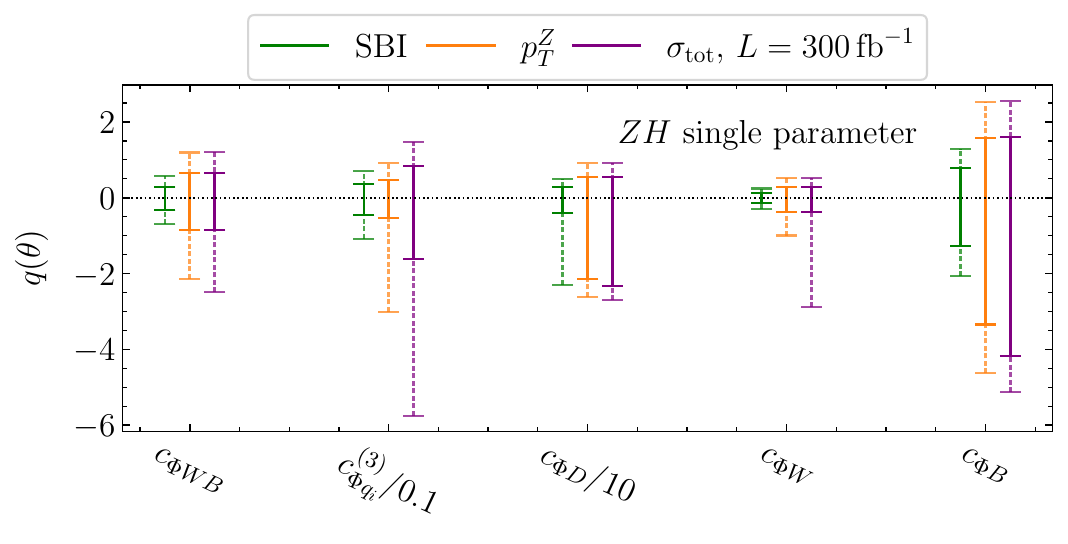}
    \caption{Projected confidence intervals for the individual processes. The small horizontal lines indicate the one- and two-sigma confidence intervals.}
    \label{fig:limits_projected_inidivual}
\end{figure}
%----------------------------

%----------------------------
\begin{figure}[t]
    \centering
    \includegraphics[width=0.5\linewidth]{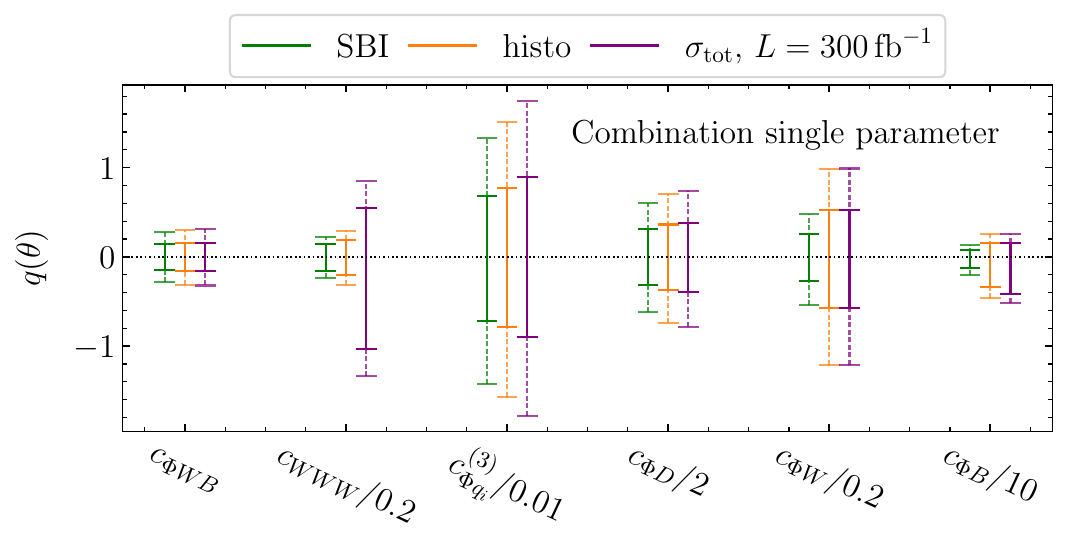}
    \caption{Projected confidence intervals for the combined likelihood (lower). The small horizontal lines indicate the one- and two-sigma confidence intervals.}
    \label{fig:limits_projected_combined}
\end{figure}
%----------------------------

We show the one-dimensional limits for single Wilson coefficients setting all other Wilson coefficients to zero Figs.~\ref{fig:limits_projected_inidivual} and~\ref{fig:limits_projected_combined}. Compared to the profiled limits, the difference between the SBI and histogram-based limits is smaller, since only a single Wilson coefficient needs to be constrained. SBI, however, still provides better or on-par sensitivity.

%%%%%%%%%%%%%%%%%%%%%%%%%%%%%%%%%%%%%%%%%%%%%%%%%%
\clearpage
\bibliography{bibliography}

\end{document}

%% file: incl_settings.tex
\usepackage[utf8]{inputenc} % input encodings (allow UTF-8 input)
\usepackage[T1]{fontenc} 	% selecting font encodings (use 8-bit T1 fonts)
\usepackage[english]{babel} % multilingual support (English language/hyphenation)

\usepackage[bitstream-charter]{mathdesign}
\usepackage{geometry} 		% flexible and complete interface to document dimensions
\usepackage{amsmath} 		% math package (American Mathematical Society)
\usepackage{mathtools} 		% math package (fixes various deficiencies of amsmath)
\usepackage{float} 			% floating objects such as figures and tables
\usepackage{graphicx} 		% enhanced support for graphics
\usepackage{tabularx} 		% tabulars with adjustable-width columns
\usepackage{booktabs} 		% professional-quality tables
\usepackage{color, xcolor} 	% foreground and background colour management
\usepackage{pdfpages} 		% inclusion of external multi-page PDF documents
\usepackage{extarrows} 		% extra arrows beyond those provided in amsmath
\usepackage{multirow} 		% create tabular cells spanning multiple rows
\usepackage{multicol} 		% intermix single and multiple columns
\usepackage{enumitem} 		% control layout of itemize, enumerate, description
\usepackage{xspace} 		% define commands that appear not to eat spaces
\usepackage{stackrel} 		% enhancement to the \stackrel command
\usepackage{tikz} 			% create PostScript and PDF graphics
\usepackage{braket} 		% Dirac bra-ket notation
\usepackage{bm} 			% access bold symbols in maths mode
\usepackage{tensor} 		% typeset tensors (tensor-style super- and subscripts)
\usepackage{slashed} 		% slash through characters (Feynman slash notation)
\usepackage{siunitx} 		% SI units package (typesetting values with units)
\usepackage{lastpage} 		% reference last page
\usepackage{cite} 			% improved citation handling
\usepackage[normalem]{ulem} % package for underlining
\usepackage{fontawesome} 	% access to web-related icons
\usepackage{tocloft} 		% control over the typography of the TOC, LOF, and LOT
\usepackage{titlesec} 		% interface to sectioning commands (various title styles)
\usepackage{doi} 			% create correct hyperlinks for DOI numbers
\usepackage{hyperref} 		% hypertext links (handel cross-referencing commands)
\usepackage[most]{tcolorbox} 					% coloured and framed text boxes
\usepackage[nameinlink, capitalize]{cleveref} 	% intelligent cross-referencing
\usepackage[nottoc, notlot, notlof]{tocbibind} 	% add references/index/contents to TOC
\usepackage[ruled, vlined]{algorithm2e} 		% floating algorithm environment
\usepackage{makecell}
\usepackage[makeroom]{cancel}
\usepackage{feynmf}

\makeatletter
\def\BState{\State\hskip-\ALG@thistlm}
\makeatother

\makeatletter
\@ifundefined{pdfoutput}{}{\DeclareGraphicsRule{*}{mps}{*}{}}
\makeatother

\makeatletter
\DeclareRobustCommand*{\bfseries}{%
   \not@math@alphabet\bfseries\mathbf
   \fontseries\bfdefault\selectfont
   \boldmath
}
\makeatother

\hypersetup{
	pdftitle={Unbinning global LHC analyses},
	pdfauthor={Bahl et al.},
	colorlinks=true, 			% false: boxed links, true: colored links
	linkcolor={red!50!black}, 	% color of internal links (sections, pages, etc.)
	citecolor={blue!50!black}, 	% color of citation links (links to bibliography)
	urlcolor={blue!80!black} 	% color of URL links (external links)
} 
\DeclareSymbolFont{usualmathcal}{OMS}{cmsy}{m}{n}
\DeclareSymbolFontAlphabet{\mathcal}{usualmathcal}

\SetArgSty{textnormal}
\SetKwComment{Comment}{{\small\#}~}{}
\SetCommentSty{mycommfont}

\setitemize{itemsep=0pt, parsep=0pt} 				% adjust itemize environment
\setenumerate{itemsep=0pt, parsep=0pt} 				% adjust enumerate environment
\newlist{todolist}{itemize}{2}
\setlist[todolist]{label=$\square$}
\usepackage{pifont}
\setitemize{itemsep=2pt,topsep=2pt,parsep=0pt,partopsep=0pt,leftmargin=*}
\setenumerate{itemsep=0pt,topsep=2pt,parsep=0pt,partopsep=0pt,labelindent=3pt,leftmargin=*}
\usepackage{amsmath}
 
\usepackage{amsthm} 		% math package (typesetting theorems using AMS style)
\theoremstyle{definition}

%% file: incl_shortcuts.tex
\definecolor{red_cb}{HTML}{e41a1c}
\definecolor{blue_cb}{HTML}{377eb8}
\definecolor{green_cb}{HTML}{4daf4a}
\definecolor{purple_cb}{HTML}{984ea3}
\definecolor{orange_cb}{HTML}{ff7f00}

\definecolor{EmeraldGreen}{HTML}{1ea78d}
\definecolor{EnglishRed}{HTML}{b02427}
\hypersetup{colorlinks=true,urlcolor=EmeraldGreen,citecolor=EmeraldGreen,linkcolor=EnglishRed}

\newcommand{\Langle}{\big\langle}
\newcommand{\Rangle}{\big\rangle}

\newcommand{\XXLangle}{\Bigg\langle}
\newcommand{\XXRangle}{\Bigg\rangle}

\def\d{\mathrm{d}}

\newcommand\one{\leavevmode\hbox{\small1\normalsize\kern-.33em1}}
\newcommand{\mean}[1]{\left\langle#1\right\rangle}

\newcommand{\loss}{\mathcal{L}} 	% loss value
\newcommand{\lag}{\mathscr{L}}

\newcommand{\madgraph}{\textsc{MadGraph5\_aMC@NLO}\xspace}

\newcommand{\pythia}{\textsc{Pythia8}\xspace}
\newcommand{\fastjet}{\textsc{FastJet}\xspace}

\newcommand{\delphes}{\textsc{Delphes}\xspace}
\newcommand{\pdfset}{\texttt{PDF4LHC15\_nlo\_30}\xspace}

\newcommand{\lhapdf}{\textsc{LHAPDF6}\xspace}

\newcommand{\arXiv}[2][]{%
	\ifthenelse{\equal{#1}{}}%
	{\href{http://arxiv.org/abs/#2}{arXiv:#2}}%
	{\href{http://arxiv.org/abs/#2}{arXiv:#2~[#1]}}}

\newcommand{\gev}{\text{GeV}}
\newcommand{\tev}{\text{TeV}}

\newcommand{\sig}{\text{sig}}
\newcommand{\bkg}{\text{bkg}}

\def\slashchar#1{\setbox0=\hbox{$#1$}           % set a box for #1
   \dimen0=\wd0                                 % and get its size
   \setbox1=\hbox{/} \dimen1=\wd1               % get size of /
   \ifdim\dimen0>\dimen1                        % #1 is bigger
      \rlap{\hbox to \dimen0{\hfil/\hfil}}      % so center / in box
      #1                                        % and print #1
   \else                                        % / is bigger
      \rlap{\hbox to \dimen1{\hfil$#1$\hfil}}   % so center #1
      /                                         % and print /
   \fi}

\newcommand{\tikznode}[2]{%
\ifmmode%
\tikz[remember picture,baseline=(#1.base),inner sep=0pt] \node (#1) {$#2$};%
\else
\tikz[remember picture,baseline=(#1.base),inner sep=0pt] \node (#1) {#2};%
\fi}

\def\mathswitchr#1{\relax\ifmmode{\mathrm{#1}}\else$\mathrm{#1}$\xspace\fi}
\def\mathswitch#1{\relax\ifmmode#1\else$#1$\xspace\fi}

\newcommand{\mat}{\mathcal{M}}

\newcommand{\ETmiss}{\ensuremath{E_{T,\text{miss}}}\xspace}

\newcommand{\OHD}{\ensuremath{\mathcal{O}_{\Phi D}}\xspace}

\newcommand{\OHW}{\ensuremath{\mathcal{O}_{\Phi W}}\xspace}
\newcommand{\OHB}{\ensuremath{\mathcal{O}_{\Phi B}}\xspace}
\newcommand{\OHWB}{\ensuremath{\mathcal{O}_{\Phi WB}}\xspace}
\newcommand{\OWWW}{\ensuremath{\mathcal{O}_{WWW}}\xspace}
\newcommand{\OHq}{\ensuremath{\mathcal{O}_{\Phi q}^{(3)}}\xspace}

\newcommand{\cHD}{\ensuremath{c_{\Phi D}}\xspace}

\newcommand{\cHW}{\ensuremath{c_{\Phi W}}\xspace}
\newcommand{\cHB}{\ensuremath{c_{\Phi B}}\xspace}
\newcommand{\cHWB}{\ensuremath{c_{\Phi WB}}\xspace}
\newcommand{\cWWW}{\ensuremath{c_{WWW}}\xspace}
\newcommand{\cHq}{\ensuremath{c_{\Phi q}^{(3)}}\xspace}